\documentclass[twocolumn]{aastex701}

\usepackage{tablefootnote}

\usepackage{scalerel}

\def\Msun{\hbox{M$_\odot$}}

\def\cm3{\hbox{cm$^{-3}$}}
\def\s1{\hbox{s$^{-1}$}}

\def\one{\,{\sc i}}             
\def\two{\,{\sc ii}}
\def\three{\,{\sc iii}}
\def\four{\,{\sc iv}}
\def\five{\,{\sc v}}
\def\six{\,\textsc{vi}}

\begin{document}

\title{Spectroscopic characterization of Young Stellar Populations and their Feedback in NGC 5253}

\author[0000-0003-4857-8699]{Svea Hernandez}
\affiliation{AURA for ESA, Space Telescope Science Institute, 3700 San Martin Drive, Baltimore, MD 21218, USA}
\email[show]{sveash@stsci.edu}

\author[0000-0002-0806-168X]{Linda J. Smith}
\affiliation{Space Telescope Science Institute, 3700 San Martin Drive, Baltimore, MD 21218, USA}
\email{lsmith@stsci.edu}

\author[0000-0002-2764-6069]{Valentina Abril-Melgarejo}
\affiliation{Space Telescope Science Institute, 3700 San Martin Drive, Baltimore, MD 21218, USA}
\affiliation{LUX, Observatoire de Paris, Universit\'{e} PSL, CNRS, Sorbonne Universit\'{e}, Meudon, 92190, France}
\email{valentina.abril@lam.fr}

\author[0000-0003-4372-2006]{Bethan James}
\affiliation{AURA for ESA, Space Telescope Science Institute, 3700 San Martin Drive, Baltimore, MD 21218, USA}
\email{bjames@stsci.edu}



\begin{abstract}
We present the spectroscopic analysis of FUV observations taken with the Hubble Space Telescope (HST) Cosmic Origins Spectrograph (COS) targeting young massive clusters in the nearby, metal-poor, blue compact dwarf \added{galaxy} NGC 5253. We characterize the stellar populations observed across seven COS pointings and report on their inferred physical parameters, age, metallicity, mass, and reddening values. Comparison between our spectroscopic ages and those inferred using photometric methods show that the former are preferentially younger. We also investigate the impact of these young massive clusters on their surrounding ISM. Using Very Large Telescope/MUSE optical observations and matching the size of the COS aperture, we measured outflow velocities of the ionized gas along the line of sight of the COS pointing with values ranging from $\sim$125-300 km s$^{-1}$. We report on strong statistically-significant correlations between the outflow velocities and the stellar ages, masses, and total mechanical luminosity primarily driven by supernovae (SNe) as derived from our full-spectrum fitting analysis. Although theoretical models predict a delayed injection of mechanical energy and momentum in low-metallicity environments ($<$0.4 Z$_{\odot}$), our study shows that in this particular system (Z$\sim$ 0.3 Z$_{\odot}$), feedback from SNe appears evident at the nominal $>$ 5 Myr ages, with no apparent delay. One possible explanation is that the decrease and suppression of mechanical feedback due to SNe explosions might be dominant at even lower metallicities than those observed in NGC 5253.

\end{abstract}



\section{Introduction} \label{sec:intro}
 The characterization of the galaxies populating the high-redshift Universe is vital to fully understand the physical and chemical conditions leading to the subsequent evolution of the cosmos. For the last several decades, blue compact dwarf galaxies \citep[BCD;][]{sar70} have gained increasing interest among the astronomical community due to their extreme ongoing star-forming activity and their quasi-pristine heavy element abundances \citep[0.02 $\leq$ Z/Z$_{\odot}$ $\leq$ 0.50;][]{kun00, amr24}, commonly forming super star clusters  \citep[SSCs; ][]{meu95, ada10}. These low-metallicity systems provide us with a unique link to the high-redshift Universe. In spite of the differences in environment between those expected in the early Universe \citep[e.g., ionizing UV background;][]{ike86,per23} and those observed in the local Universe, detailed studies of these unique systems provide critical insights into the star-formation processes in nearly pristine environments at exquisite signal-to-noise (S/N) and spatial resolution, unachievable at high-redshift. Nearby \citep[$\sim$1-200 Mpc;][]{mad13} star-forming dwarf galaxies are then the closest we can get to performing spatially-resolved ($\lesssim$ sub-kpc) studies under conditions reminiscent to those at high-redshift.\par
 When it comes to studying star-forming environments in detail, specifically their young and massive stellar populations along with their intervening interstellar medium (ISM), the far-ultraviolet (FUV) wavelength regime provides an unparalleled diagnostic power. FUV spectroscopic observations of star-forming galaxies contain a plethora of features from the hot and massive stars being created, as well as signatures of different ionization species arising from the ISM \citep{hec97, hec98,vaz04,lei11, berg22}. To add to this, the importance of the rest-frame UV continues to expand as it provides a window to the high-redshift universe through some of the largest and most sensitive ground- and space-based telescopes working in the visible and infrared. \par

The intensely star-forming, low metallicity BCD galaxy NGC~5253 has been the subject of extensive FUV spectroscopic observations with the Hubble Space Telescope (HST). \added{Different aspects of NGC 5253, including its stellar mass \citep[$\sim$11 $\times$ 10$^{8}$ M$_{\odot}$; ][]{lop12}, compact star formation \citep[starburst diameter $<$ 500 pc; ][]{wes13}, and porous ISM with low density gas \citep[as suggested by its high {[\ion{O}{3}]} 88$\mu$m/{[\ion{C}{2}]}158$\mu$m$\sim$2; ][]{cor15}, are reminiscent of the properties of distant galaxies \citep[$z\gtrsim$ 5; M$_{*}\lesssim$10$^{8}$ M$_{\odot}$, starburst diameter $\sim$ 400-1000 pc, {[\ion{O}{3}]} 88$\mu$m/{[\ion{C}{2}]}158$\mu$m$\sim$2-10; ][]{izo21, car20, wit22, mor24}. Additionally, NGC 5253} is nearby \citep[$3.32\pm0.25$~Mpc;] []{sabbi18}, has a low $12 + \log {\rm O/H}$ abundance of 8.26 (\citealt{monreal12}; or a metallicity of 
37\% solar using the O/H abundances by \citealt{asplund09}), and it is hosting an on-going starburst in the central 250--300 pc \added{with several massive star clusters} \citep[e.g.,][]{rie88, bec96, cal97, kob97, tre01,har04, deg13, cal15a}.
\par
Early studies by \citet{bec96} showed that at the center of this galaxy the starburst has to be very young as any non-thermal radio emission is very weak, suggesting that few supernovae have so far occurred. Instead they find strong thermal radio emission from a massive ultracompact H\two\ region or ``supernebula'' ionized by a very young SSC. Using HST/NICMOS imaging \citet{alonso04} reported for the first time the presence of a double nuclear star cluster (separated by $0''.3-0''.4$). A more recent study by  \citet{smith20} using HST observations mapped to the GAIA astrometric reference frame showed that NGC 5253 is instead hosting three young nuclear SSCs.
\par

\added{An} extensive photometric study by \citet{cal15a} used 13 HST filters covering the FUV to 2\,$\mu$m reporting on the 11 brightest clusters in the central 300-pc region of NGC~5253. They find that all of the clusters in their sample are younger than 15~Myr and the two nuclear clusters found by \citet{alonso04}, which they label as NGC 5253-5 and NGC 5253-11, have ages of $1\pm1$ Myr. These very young ages are in apparent contradiction with the detection of Wolf-Rayet (WR) features in cluster NGC 5253-5, implying an age of 3--5 Myr \citep[e.g.,][]{kob97, mon10}. To investigate this age discrepancy, \citet{smith16} examined STIS FUV spectroscopy of NGC 5253-5 and showed that the WR emission is instead produced by very massive stars ($>$100 \Msun) at an age of 1--2 Myr.
\par
When it comes to understanding the origin of the star formation in NGC 5253, studies have proposed that a possible encounter with the neighboring spiral galaxy, M 83, could have been the initial trigger, with subsequent episodes of elevated star formation \citep{cal89}. Such an interaction between NGC 5253 and M 83 is potentially supported by the tidal extension of \ion{H}{1} gas, observed North of NGC 5253, in the direction of M 83. \citet{lop12} studied the kinematics of the \ion{H}{1} gas in NGC 5253, and reported the detection of low-metallicity \ion{H}{1} gas in-falling onto NGC 5253, creating the observed starburst. Given the active star-formation environment (SFR = $0.1-0.36$ M$_\odot$ yr$^{-1}$; \citealt{cal04, cal15a, dal09, tur04}), studies of the ISM show extensive evidence of feedback through structures such as shells and filaments \citep{mon10,wes13, abr24}. These studies are already hinting at the critical role played by massive stars and star clusters in the starburst shaping the ISM.
\par
Another intriguing aspect of the starburst region is the presence of \added{localized nitrogen-enriched regions, with abundances higher by a factor of 2--3 than what is observed in the rest of the body of the galaxy} \citep{walsh87, kob97, lopez07,monreal10}. Understanding the source of this enrichment is highly relevant to JWST observations of \added{N-rich high-redshift low-metallicity systems which have  been observed to reach $\log$(N/O) values of $\gtrsim-$0.5-0.7}  \citep{bunker23,cameron23,schaerer24}, {compared to the more nominal values of $\log$(N/O)$\lesssim-$1.25 for galaxies with 12+$\log$(O/H)$\lesssim$ 8.5 \citep{est14, izo23}}. Very recently, \citet{pru25} studied the N enrichment in the central region of NGC 5253 at 2.3-pc resolution with MUSE-NFM \added{and found that the peak of the enrichment is located north of the nuclear clusters and is consistent with their stellar wind feedback. Additionally, they report that the regions with the most extreme enhancements reach values of $\log$(N/O)=$-$0.65 reminiscent of the enhancements being observed at high redshift, compared to the median value of $\log$(N/O)$\sim-$1.25 for the rest of the galaxy, the latter value aligning well with observations of local star-forming galaxies.}
\par
In this work we present a spectroscopic characterization of the young stellar clusters in NGC 5253. Our analysis is based on archival HST/COS FUV observations of seven pointings targeting the brightest UV objects in this star-forming galaxy. In an effort to better understand the possible impact of these massive star clusters and their feedback onto the surrounding ISM, we also analyzed spatially-matched VLT/MUSE spectra to quantify the outflow velocities of the ionized gas. This paper is structured as follows: in Section \ref{sec:obs} we provide a concise description of the HST/COS archival observations along with the data processing steps; in Section \ref{sec:stellar} we present an overview of typical FUV stellar features of young populations; in Section \ref{sec:ana} we detail the adopted analysis to characterize the stellar populations and \added{the kinematics of the surrounding ionized gas}; and in Sections \ref{sec:discussion} and \ref{sec:sum} we present our primary discussion points and concluding remarks, respectively. 
 \begin{figure}[ht]
   	  \center{\includegraphics[scale=0.45]{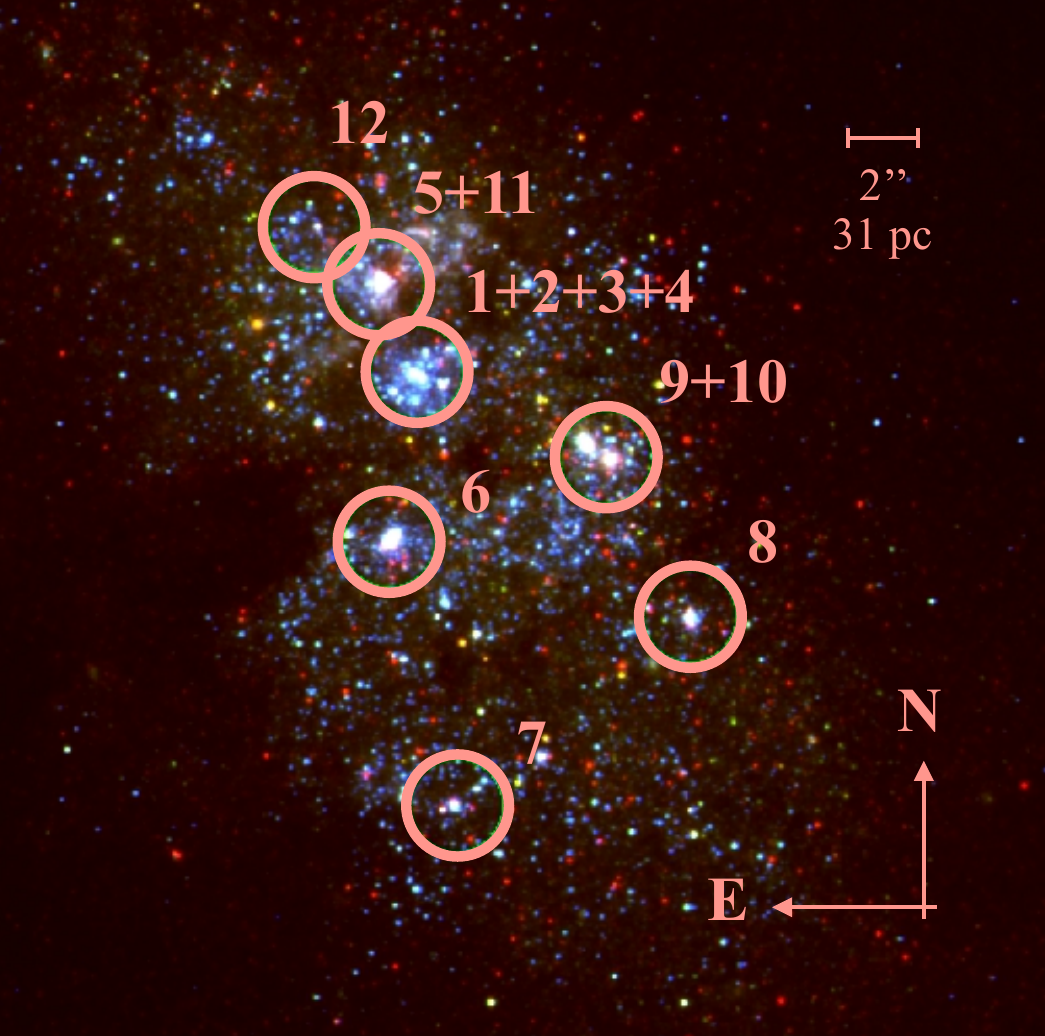}}
      \caption{HST color composite of the blue compact dwarf NGC 5253 \citep[LEGUS;][]{cal15b}. Blue: HST/WFC3/F275W; Green: HST/ACS/F435W; Red: HST/ACS/F814W. The COS pointings are shown with $2''.5$ circular apertures.}
         \label{fig:pointings}
   \end{figure}

\section{Observations and Data Reduction} \label{sec:obs}
The analysis presented as part of this work relied on archival observations taken as part of HST program IDs 11579, 15193 (PI: Aloisi), 16045 (PI: Smith), and 16240 (PI: James). These programs collected COS observations for a total of 7 pointings, targeting young massive clusters (YMCs) throughout NGC 5253. For three of these targets the programs collected data using the G130M and G160M settings providing wavelength coverage between 1150 to 1775 \r{A}. The rest of the targets were observed with the G130M grating alone, with wavelength coverage of 1100 to 1425 \r{A}. In Table \ref{tab:cos_obs} we list the exposure information, including program ID, dataset name, observation date, grating, central wavelength, exposure time, and S/N ratios. \par

Figure ~\ref{fig:pointings} shows the central region of NGC 5253 as observed in HST images from LEGUS \citep{cal15b}. This figure highlights that the 2\arcsec.5 COS aperture encompasses in several pointings more than a single bright cluster, e.g., NGC 5253-9+10. We note that the majority of the clusters in this study are identified with the nomenclature from \citet{cal15a}, with an additional cluster from HST program ID 16045, which is labeled as NGC 5253-12.

\par
The data were downloaded from the Mikulski Archive for Space Telescopes (MAST) and calibrated with the HST pipeline, \texttt{CalCOS} V3.4.7 \citep{sod22}. To obtain a single co-added spectrum per target, we weight-combined the individual x1d exposures using the \texttt{IDL} software developed by the COS Guaranteed Time Observed team \citep{dan10}. The individually-extracted spectra were combined weighting by exposure time and taking into account the data quality information flagged by \texttt{CalCOS}. As a final step in the data reduction, we binned the spectra by 6 pixels (1 resolution element). The S/N values per resolution element of our sample range between 12 and 23 (estimated at $\lambda$= 1270 \r{A}, last column in Table \ref{tab:cos_obs}). We assessed the wavelength accuracy of our coadded observations by comparing the laboratory wavelengths from Milky Way (MW) absorption lines (i.e., \ion{Si}{2} 1190.42 \r{A}) to the empirical values as measured from the COS spectra. Overall, we find on average offsets of the order of 0.03 \r{A}, well within the required wavelength accuracy of the instrument \citep[0.06 \r{A}; ][]{sod22}. The {\it HST} data used in this paper can be found in MAST: \dataset[10.17909/fd6e-an91]{https://doi.org/10.17909/fd6e-an91}.\par
    \movetabledown=1.2in	
	\begin{rotatetable}
         \begin{deluxetable*}{cccccccccc}
     \tablecaption{COS Exposure Information\label{tab:cos_obs}}
    \tablehead{  \colhead{Target} &  \colhead{Program}  &  \colhead{R.A.} &  \colhead{Dec.}&  \colhead{Dataset} &  \colhead{Obs. Date}  &  \colhead{Grating} &  \colhead{CENWAVE} &  \colhead{Exp. Time} &  \colhead{S/N$^{a}$} \\
    \colhead{} &  \colhead{ID}  &  \colhead{(J2000)} &  \colhead{(J2000)} &  \colhead{} &  \colhead{(YYYY-MM-DD)}  &  \colhead{(\AA)} & \colhead{}&  \colhead{(sec)} &  \colhead{}}
\startdata
NGC 5253-1+2+3+4 &	16240& 13 39 55.91 & $-$31 38 27.06	&LEC70020Q& 	2021-05-11 &	G130M &	1291	&2240 & 15\\
	&&&&	LEC701030 &	2021-05-11 &	G130M &	1222	&1782 \\
NGC 5253-5+11 &	16240	& 13 39 55.98 & $-$31 38 24.59	&LEC702020 	&2021-05-12 &	G130M &	1291	&7624 & 23 \\
	&&&&	LEC703020 &	2021-03-01 &	G130M &	1222&	4873 \\
	&&&&	LEC703030 &	2021-03-01 &	G130M 	&1291	&2176 \\
	&&&&	LEC706020 &	2023-05-08& 	G130M &	1291	&7619 \\
 NGC 5253-6 & 	11579	& 13 39 56.02 & $-$31 38 31.30	&LB7H61010 &	2010-07-03 &	G130M &	1291&	3038 & 13\\
	&15193	&&&LDN706030 &	2019-05-03 &	G130M &	1222	&1655 \\
	&	&&&LDN706040 &	2019-05-03 &	G160M 	&1623	&3472 \\
NGC 5253-7	&11579	& 13 39 55.89 & $-$31 38 38.34	&LB7H61020 &	2010-07-03 	&G130M& 	1291&	6295 & 19\\
&	15193&&&	LDN756010 &	2019-08-14 &	G130M &	1222	&3012 \\
	&&&&	LDN706010 	&2019-05-03 &	G160M &	1623	&3408 \\
&&&&		LDN756020 &	2019-08-14 &	G160M &	1623&	3924 \\
	&&&&	LDN706020 &	2019-05-03 	&G130M &	1222	&3012 \\
NGC 5253-8&	16240	& 13 39 55.35 & $-$31 38 33.40	&LEC704010&	2021-02-22	&G130M	&1291	&4936 &13\\
&		&  & 	&LEC704020&	2021-02-22	&G130M	&1291	&5236 &\\
NGC 5253-9+10 	&16240& 13 39 55.54 & $-$31 38 29.05		&LEC705020 	&2021-05-11 &	G130M &	1291&	3595 &12\\
	&&&&	LEC705030 	&2021-05-11 	&G130M 	&1222&	3215 \\
NGC 5253-12&	16045&	 13 39 56.11 & $-$31 38 23.22	&LE8901020&	2020-02-10&	G130M	&1291	&10239 & 15\\
&	&	  & 	&LE8902020&	2020-02-10&	G160M	&1600	&10337 & \\
\enddata
\begin{minipage}{20cm}~\\\\\\
 \footnotesize{$^{a}$ S/N values per resolution element estimated from the coadded products at 1270 \r{A}.}
\end{minipage} 
\end{deluxetable*}
\end{rotatetable}
\clearpage
\begin{figure*}
  \center{\includegraphics[scale=0.3]{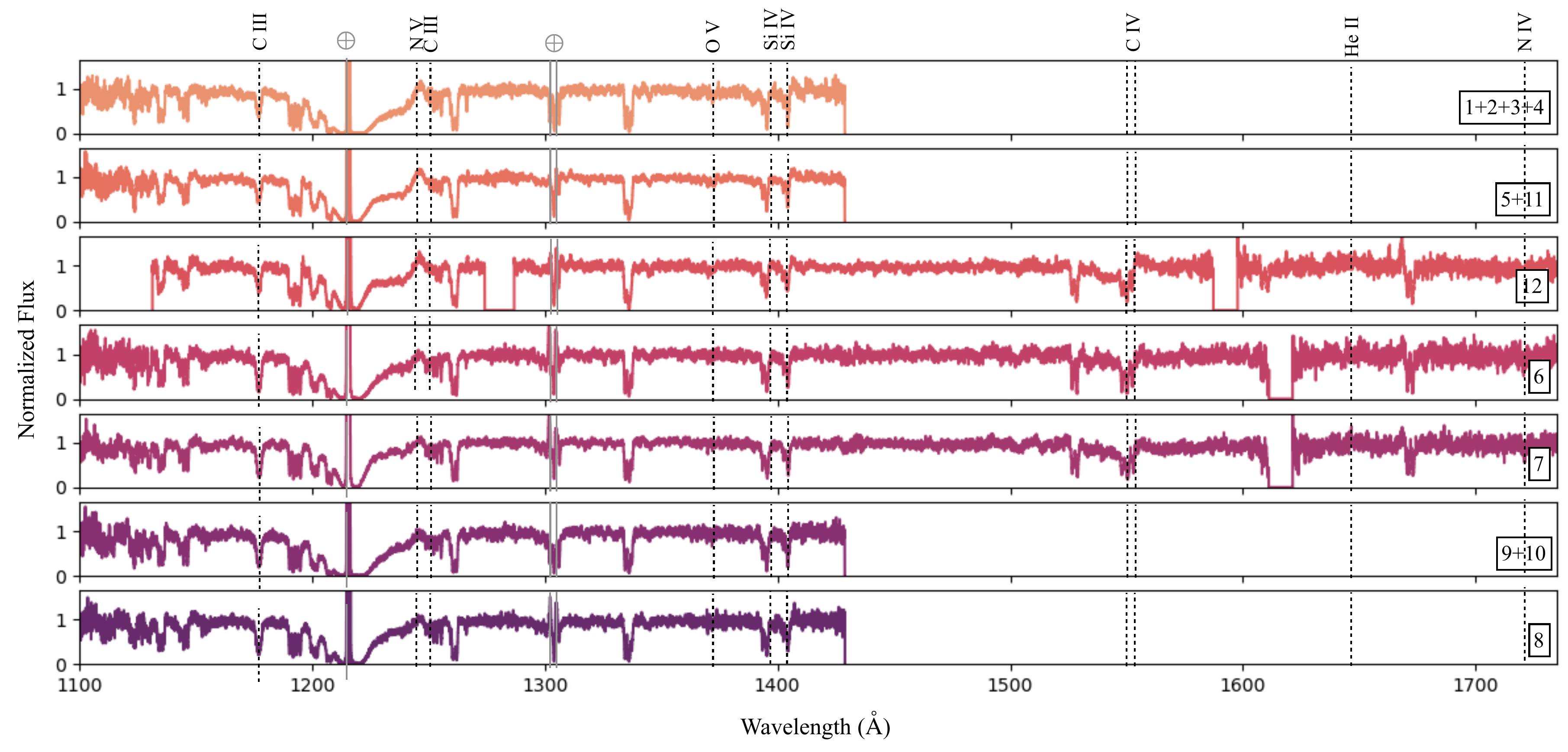}}
\caption{Normalized COS spectra for the pointings in NGC 5253. The data are binned by a COS resolution element (6 pixels). We show with dashed vertical lines photospheric and stellar wind lines. Solid gray vertical lines show the location of contaminating geocoronal emission, $\bigoplus$. From top to bottom the spectra are ordered in decreasing strength of the \ion{N}{5} feature, which is strongest at stellar ages $<$3 Myr \citep{chi19}.} 
    \label{fig:cos_spectra}
\end{figure*}
\section{Stellar Features in Young Stellar Populations} \label{sec:stellar}

The properties of stellar populations are commonly inferred by fitting spectral energy distributions (SEDs) to photometric measurements covering a broad wavelength range \citep[e.g., ][]{cal15a,ada17, tur21}. This technique, however, is challenged by the known age/metallicity/reddening degeneracy problem \citep[][]{whi20,whi23}. These issues can be circumvented by fitting key stellar spectral features to theoretical stellar templates. This approach simultaneously determines intrinsic reddening, and the stellar ages and metallicities of these same populations \citep{chi19}. \par

In Figure \ref{fig:cos_spectra} we show the normalized COS spectra, ordered in decreasing strength of the \ion{N}{5} $\lambda\lambda$1238,1242 spectral features. The most prominent stellar features in our sample are strong P-Cygni profiles (broad blueshifted absorption combined with redshifted emission) from different ions. These P-Cygni features arise from strong winds that are radiatively driven off the photospheres of the young massive stars present in the star clusters. The terminal velocities of these winds are traced by the observed blue edge velocities of the blueshifted absorption component \citep[e.g.,][]{lei11}. Specifically the \ion{N}{5} $\lambda\lambda$1238,1242, \ion{Si}{4} $\lambda\lambda$1393,1402, and \ion{C}{4} $\lambda\lambda$1548,1550 line profiles are generated by O and B stars, and are both metallicity and age sensitive. In our sample we detect all of these transitions at different strengths. \ion{N}{5} (ionization potential, \added{IP=77.47} eV) is one of the FUV doublets with the highest ionization state, and is strongest in stars with stellar ages of $<$3 Myr \citep{chi19}. We note that pointings NGC 5253-1+2+3+4, NGC 5253-5+11, and NGC 5253-12 in our sample show the strongest \ion{N}{5} profiles. \ion{O}{5} $\lambda$1371 (IP= 77.41 eV), also a high ionization line, traces the outflowing matter from the hottest massive stars. In contrast to \ion{N}{5} whose strength weakens at older ages, the \ion{O}{5} P-Cygni profile fully disappears as soon as the stellar populations reach ages $>$2.5 Myr and the hottest stars have evolved off the main sequence \citep{smith16,smi23, wof23}. Three of the NGC 5253 pointings (1+2+3+4, 5+11, 12) show hints of weak P-Cygni \ion{O}{5} features. We also highlight that although present in the NGC 5253 spectra, the \ion{Si}{4} $\lambda\lambda$1393,1402 and \ion{C}{4} $\lambda\lambda$1548,1550 wind lines are contaminated by strong \ion{Si}{4} and \ion{C}{4} ISM absorption arising from the interstellar gas in NGC 5253 \citep{abr24}. \par
Aside from the strong P-Cygni lines, the COS data reveal other well-developed and strong photospheric features, such as the \ion{C}{3} $\lambda$$\lambda$1175,1247 lines. In general, these photospheric lines are weaker than the stellar wind profiles, and typically require high S/N spectra and medium spectral resolution for detection. According to \citet{pel02} the strength of these \ion{C}{3} lines increases as the temperature of these O and B stars decreases. We detect strong absorption of \ion{C}{3} lines in all of the COS pointings (see blue end of the spectra in Figure \ref{fig:cos_spectra}). 
 \begin{figure*}[!ht]
   	  \centerline{\includegraphics[scale=0.6]{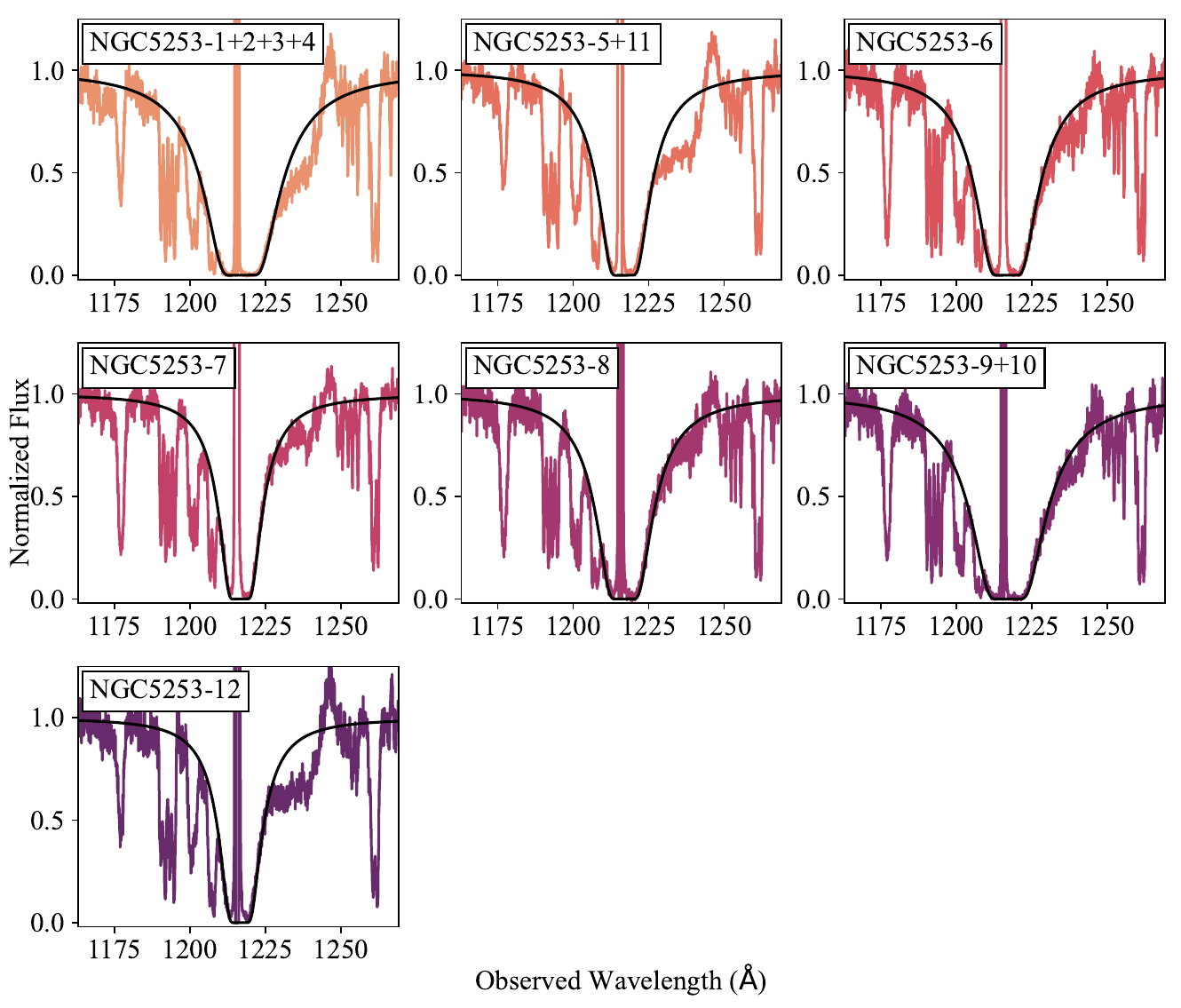}}
      \caption{Lyman $\alpha$ absorption in the COS observations (colored). In black we show the best fitting Voigt models which include both the Milky Way and the NGC 5253 components. The cluster pointings are shown in each subpanel.}
         \label{fig:HI_mod}
   \end{figure*}
\begin{table}[!ht] 
  \centering
    \caption{\ion{H}{1} column densities}\label{tab:HI}
    \def\arraystretch{1.25}
     \begin{tabular}{l c }
    \hline
    \hline
   Target & log$N$(\ion{H}{1})  \\
   & (cm$^{-2}$)\\
    \hline
    \hline
NGC 5253-1+2+3+4& 	21.48 $\pm$ 0.02\\
NGC 5253-5+11 &	21.08 $\pm$	0.08\\
NGC 5253-6&	21.29	$\pm$0.23\\
NGC 5253-7&	20.83$\pm$	0.37\\
NGC 5253-8&	21.18	$\pm$0.05\\
NGC 5253-9+10&	21.47	$\pm$0.04\\
NGC 5253-12&	20.80 $\pm$	0.09\\
    \hline
    \end{tabular}
\end{table} 

\section{Analysis} \label{sec:ana}
\subsection{Lyman $\alpha$ correction}\label{sec:lyman}
Before proceeding with the analysis of the stellar populations targeted in our COS observations, we first model and correct for the strong \ion{H}{1} absorption profiles in the data. At the distance of NGC 5253 \citep[3.32 Mpc;][]{sabbi18}, the Lyman $\alpha$ absorption profiles imprinted in the COS spectra are a combination of the \ion{H}{1} gas in the Milky Way, and neutral gas intrinsic to the galaxy. This correction is particularly critical as the essential \ion{N}{5} doublet \citep[age indicator;][]{chi19} is contaminated by the red wing of the Lyman $\alpha$ absorption.  \par
Similar to previous studies of nearby young stellar clusters \citep[e.g., ][]{sir22,smi23}, before fitting the Lyman $\alpha$ absorption, the COS spectra are normalized by fitting the stellar continuum. This is done by strategically placing nodes on featureless regions of the spectrum (free of ISM or stellar absorption/emission), and applying a cubic spline interpolation. We then model the two \ion{H}{1} components by fitting Voigt profiles at different radial velocities, using the Python tool \texttt{VoigtFit} \citep{kro18}. We fix the column density, log$N$(\ion{H}{1}), of the Milky Way component, adopting a value of 20.57 cm$^{-2}$ from the \ion{H}{1} map of \citet{hi416}, observed in the direction of the NGC 5253 galaxy. \par
In Figure \ref{fig:HI_mod} we show the Lyman $\alpha$ absorption features in our sample, along with the best fitting Voigt models in black. Our modeling of the column densities of \ion{H}{1}, specifically intrinsic to NGC 5253, highlight the dramatic variations in neutral gas across the galaxy. We find the lowest column densities, of the order of log$N$(\ion{H}{1}) $\sim$ 20.8 cm$^{-2}$, in the outer-most COS pointings; NGC 5253-7 in the southern regions of the galaxy, and NGC 5253-12 in the northern regions. For the rest of the pointings, we estimate slightly higher values, log$N$(\ion{H}{1}) $\sim 21.0-21.5$ cm$^{-2}$. In Table \ref{tab:HI} we list the final \ion{H}{1} column densities along with their uncertainties.\par
We compare our \ion{H}{1} column densities with those estimated in a recent study by \citet{abr24}, where they perform a multi-phase ISM analysis using the same COS observations. Overall, we find that our measured column densities agree within the inferred uncertainties, with the exception of NGC 5253-1+2+3+4, which deviates from our value by 2 $\sigma$. Assuming the values in Table \ref{tab:HI}, we produce Voigt profiles for each individual COS pointing and remove the \ion{H}{1} ISM contributions to obtain  Lyman-$\alpha$ corrected spectra. \par

\subsection{Stellar Properties with SESAMME}\label{sec:sesamme}
To characterize the dominant stellar population in each of the COS pointings, we make use of the recently released \texttt{SESAMME} software \citep{jon23}. \texttt{SESAMME} is a Python-based full spectrum fitting code that simultaneously infers the age, metallicity, extinction and, if the distance to the target is known, the stellar mass. The software samples the posterior probability distribution in these four dimensions, applying Markov Chain Monte Carlo methods. We note that \texttt{SESAMME} fits individual stellar populations for each input spectrum, which for our sample assumes a single dominant stellar cluster inside the COS apertures. The implications of this assumption are discussed in Section \ref{sec:discussion} where we compare the inferred properties with \texttt{SESAMME} against studies in the literature that resolve individual clusters.\par

As detailed in \citet{jon23}, before running \texttt{SESAMME}, we pre-processed the COS spectra to (1) remove the Lyman-$\alpha$ absorption profile as described in Section \ref{sec:lyman}, (2) correct for Galactic extinction, (3) correct for the radial velocity of the target and (4) smooth and resample to a similar resolution as that of the theoretical models. We correct for Galactic extinction adopting a color excess value of  E(B-V)$_{MW} =$0.049 from \citet{sch11} (as listed in the NASA/IPAC Extragalactic Database) along with the reddening law by \citet{gor09}. For correction (3), similar to past studies of UV star clusters \citep{sir22}, we have used the \ion{C}{3} $\lambda\lambda$1175, 1247 lines to estimate the radial velocity of the targeted clusters. In the second column of Table \ref{tab:sesamme}, we list the measured radial velocities for the different targets. \added{Lastly, to accurately compare the observations with the stellar population models we accounted for their differing spectral resolutions, step (4) above, by smoothing the COS spectra with a Gaussian filter with a $\sigma_{\rm sm}$ = $\sqrt{\sigma^{2}_{\rm model}-\sigma^{2}_{\rm COS}}$, where $\sigma_{\rm model}$ is dependent on the models used, and $\sigma_{\rm COS}= \rm FWHM_{\rm COS}/2.355\; = 0.06\: \rm \AA$/2.355. As part of this last step, the smoothed spectrum is then resampled to the wavelength grid of the model}.\par
 \begin{figure*}
   	  \centerline{\includegraphics[scale=0.5]{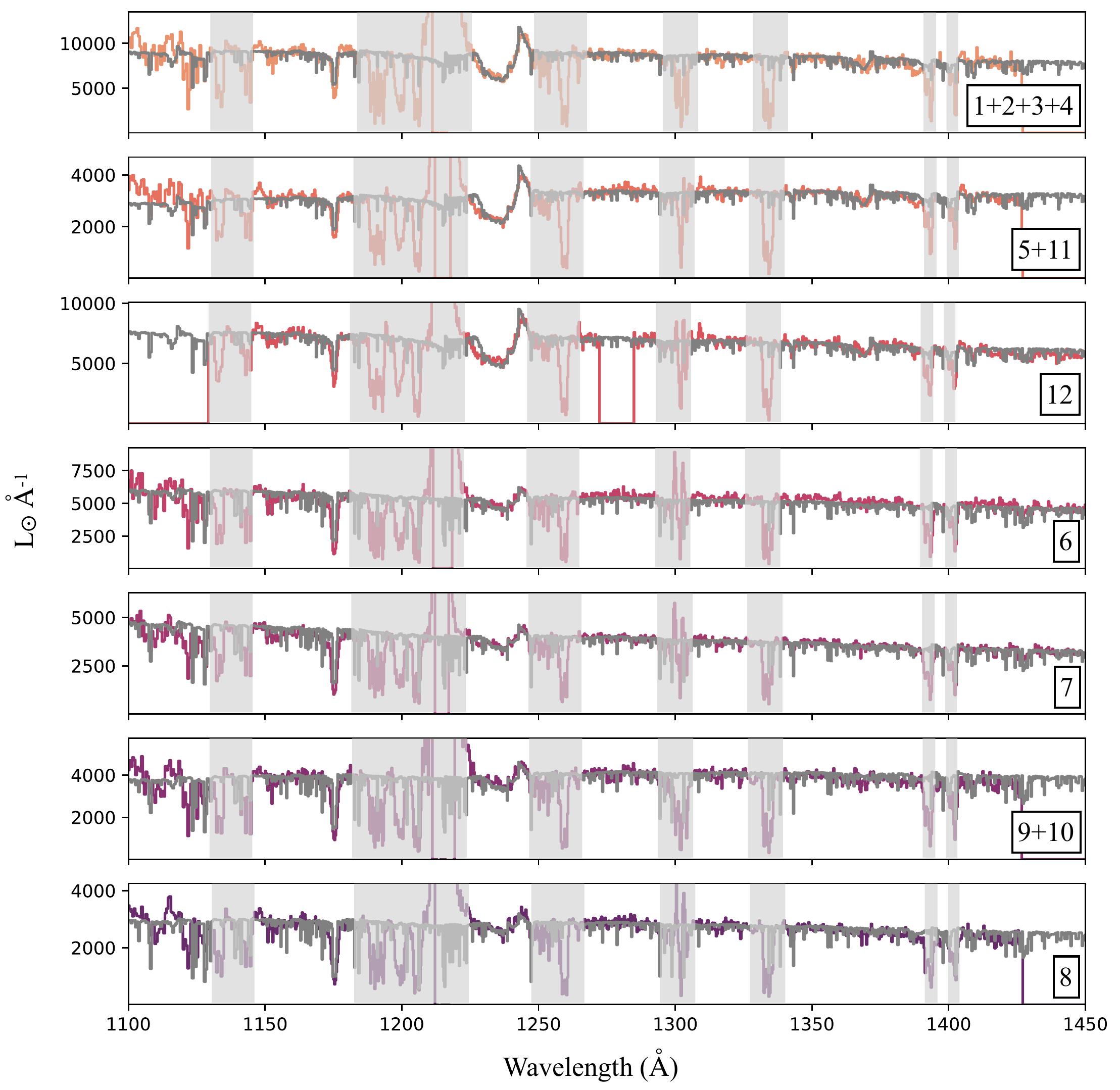}}
      \caption{Example synthesis fits for each of the COS pointings. In gray we show the best stellar model identified by \texttt{SESAMME}. We note that the model does not include ISM contributions. The gray-shaded regions show the wavelengths masked in our full spectral fitting analysis which primarily target strong absorption from the cold ISM.}
         \label{fig:mod}
   \end{figure*}

For the stellar population analysis with \texttt{SESAMME} we adopted the starburst \added{attenuation} curve by \citet{cal00} along with theoretical models of \texttt{Starburst99} v7.0 \added{\citep[][assuming $\sigma_{\rm model}= \rm FWHM_{\rm model}/2.355\;  = 0.4\: \rm \AA$/2.355]{lei99}}. 

\added{Relevant to the theoretical models, in general nebular continuum emission can account for a large fraction of the total emission along the line of sight to the targeted clusters. The contribution from this nebular continuum component is even more sizable for young and metal-poor populations \citep{sch02, and03, mir25}. The adopted \texttt{Starburst99} models account for this additional component, estimating this contribution to the integrated spectra using the wavelength-dependent emission coefficients by \citet{fer80} and \citet{all84} assuming Case B conditions and free-free, bound-free and two-photon emission from both H and He$^{+}$.}
 \begin{figure*}
   	  \centerline{\includegraphics[scale=0.45]{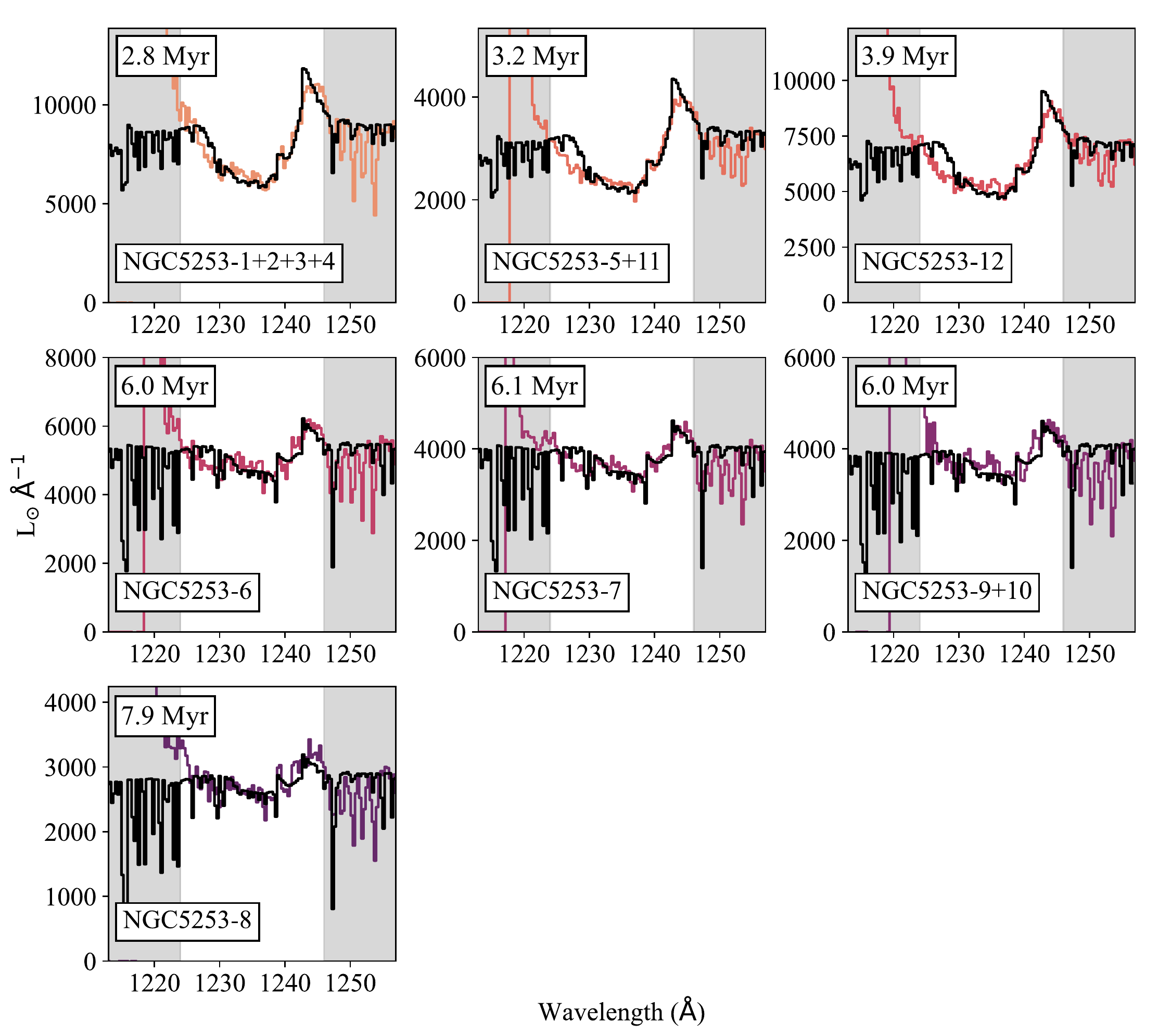}}
      \caption{Zoomed panels highlighting the agreement between the best fit model identified by \texttt{SESAMME} in black and the observed \ion{N}{5} doublet in color. The gray-shaded regions show the wavelengths masked in our analysis. We show in each panel the inferred ages, also provided in Table \ref{tab:sesamme}.} 
         \label{fig:mod_NV}
   \end{figure*}
   
\added{As part of our analysis we also made use of the \texttt{NumPy} boolean array object within \texttt{SESAMME} to fully mask problematic wavelength regions avoiding detector gaps, geocoronal and ISM contamination. More details on the specific usage of this feature in \texttt{SESAMME} can be found in \citet{jon23}. Specifically on the identification of contaminating ISM absorption, we relied on the UV spectral line list of \citet{lei11} for wavelengths $\gtrsim$1150 \AA; for bluer wavelengths not covered in this library we made use of the COS line list used in the ISM analysis by \citet{jam14}. Through visual inspection we selected the most prominent absorption and masked them as shown in the gray-shaded regions in Figures \ref{fig:mod} and \ref{fig:fits_g160m}. We note that the masked wavelength windows were designed to exclude the ISM contamination from both the MW and NGC 5253.}

\added{For the remainder of the analysis with \texttt{SESAMME}} we use the default number of walkers (128), and run the code with chains of 50,000 steps. As a last step in the analysis with \texttt{SESAMME} we initialize the walkers around values obtained by maximizing the likelihood and after completing each individual run we thin the chain by an autocorrelation time, which was in general of the order of a few hundred steps.

\subsection{SESAMME's Sensitivity to Wavelength Coverage}
As detailed in Section \ref{sec:obs}, for three COS pointings (NGC 5253-6, NGC 5253-7, NGC 5253-12) the HST programs collected observations with wavelength coverage between 1150 to 1775 \r{A} (see Figure \ref{fig:cos_spectra}). For the rest of the pointings, the observations were taken with the G130M setting alone, with wavelength coverage between $\sim$1100 and 1425 \r{A}. To test the sensitivity of \texttt{SESAMME} to wavelength coverage, we ran the code on the observations for NGC 5253-6, NGC 5253-7, NGC 5253-12 with (a) the full wavelength range, $\sim$1100-1735 \r{A}, and (b) limiting the wavelength range to $\sim$1100-1425 \r{A}. Overall, we find that a more extended wavelength range in the input spectrum to \texttt{SESAMME} has a negligible impact on the inferred age and metallicity. A limited wavelength coverage, however, seems to influence the output stellar masses and E(B-V) values, estimating both higher masses by $\lesssim$ 50\% and higher E(B-V) values by $\sim$ 30 \% than those obtained from the spectra with  broader coverage. Comparing the inferred values from these two \texttt{SESAMME} runs against those inferred by the photometric study by \citet[][using a total of 13 different filters including the critical FUV bandpass]{cal15a}, we report that the values obtained in the extended wavelength coverage runs are in better agreement than when assuming a limited wavelength coverage. We discuss in Section \ref{sec:literature_stellar} some of the possible degeneracies present in photometric studies.

\begin{table*}
  \centering
    \caption{Stellar properties from the COS sample inferred with \texttt{SESAMME}}\label{tab:sesamme}
    \def\arraystretch{1.25}
     \begin{tabular}{l cccccccc }
    \hline
    \hline
   Target & RV$^{a}$ & Age & $Z$ & M & E(B-V) & log ($Q_{\rm H}$) & log ($E_{\rm winds}$) & log ($E_{\rm Total}$)$^{b}$\\
   & (km s$^{-1}$) & (Myr) & ($Z_\odot$) &(10$^{4}$ M$_{\odot}$) &(mag) & (photons s$^{-1}$) & (erg) & (erg)\\
    \hline
    \hline
NGC 5253-1+2+3+4& 385	&2.8 $^{+0.6}_{-1.4}$ &0.27 $^{+0.10} _{-0.07}$ & 4.86 $^{+0.07} _{-0.01}$ &  0.15$^{+0.05} _{-0.01}$ & 52.7 & 53.2 &   53.2\\
NGC 5253-5+11 & 377&	3.2 $^{+0.8} _{-0.7}$&0.27 $^{+0.10} _{-0.07}$& 6.37$^{+0.02} _{-0.01}$ &0.27 $^{+0.01} _{-0.01}$ & 52.7 & 53.2 & 53.2\\
NGC 5253-6& 412&	6.0 $^{+0.4} _{-0.3}$&	0.22 $^{+0.12} _{-0.11}$	& 6.43 $^{+0.01} _{-0.01}$ &	0.14 $^{+0.01} _{-0.01}$ & 51.8 & 53.4 & 54.3\\
NGC 5253-7	&445& 6.1 $^{+0.6} _{-0.4}$&	0.28 $^{+0.10} _{-0.07}$&	3.05 $^{+0.02} _{-0.01}$ &	0.10 $^{+0.03} _{-0.01}$ & 51.8 & 53.4 & 54.3\\
NGC 5253-8& 388&	7.9 $^{+0.3} _{-0.3}$&	0.27 $^{+0.07} _{-0.10}$&	9.33 $^{+0.01} _{-0.01}$	&0.20$^{+0.01} _{-0.01}$ & 51.1 & 53.5 & 54.5\\
NGC 5253-9+10& 411&	6.0 $^{+0.4} _{-0.4}$	&0.27 $^{+0.10} _{-0.07}$	&8.04 $^{+0.01} _{-0.01}$ &	0.18 $^{+0.01} _{-0.01}$ & 51.8 & 53.4 & 54.3\\
NGC 5253-12 & 386&	3.9 $^{+0.4} _{-0.3}$&	0.27 $^{+0.10} _{-0.07}$	&2.06 $^{+0.01} _{-0.01}$ &	0.14 $^{+0.01} _{-0.01}$ & 52.3 & 53.3 & 53.8\\
						
    \hline
    \end{tabular}
       \begin{minipage}{15cm}~\\
 \footnotesize{
 $^{a}$ Values estimated from the COS observations using the \ion{C}{3} $\lambda\lambda$1175, 1247 photospheric lines.  \\
 $^{b}$ Refers to the total mechanical energy injected by stellar winds and SNe. \\

 }
\end{minipage} 
\end{table*}

\begin{table*}[!h] 
  \centering
    \caption{Sample properties from the literature as reported by \citet{cal15a}}\label{tab:properties}
    \def\arraystretch{1.25}
     \begin{tabular}{l cc cccc }
    \hline
    \hline
   Target & Age$_{\rm phot}$ & $\langle$ Age$_{\rm phot}$ $\rangle$$^{a}$ & M$_{\rm phot}$ & M$_{\rm Total}$$^{b}$ & E(B-V)$_{\rm phot}$ & $\langle$E(B-V)$_{\rm phot}$$\rangle$ $^{c}$\\
   & (Myr) & (Myr) & (10$^{4}$ M$_{\odot}$)& (10$^{4}$ M$_{\odot}$) &(mag)&(mag) \\
    \hline
    \hline
NGC 5253-1&	5 $^{+1}_{-2}$ & 5$^{+1}_{-2}$ & 1.05$^{+0.28}_{-0.22}$ & 4.04$^{+0.17}_{-0.15}$ & 0.12$^{+0.03}_{-0.03}$ & 0.14$^{+0.02}_{-0.02}$ \\
NGC 5253-2&	5 $^{+1}_{-2}$  && 0.91$^{+0.31}_{-0.22}$ & &  0.08$^{+0.03}_{-0.03}$ & \\
NGC 5253-3&	5 $^{+1}_{-0}$  &  & 0.46$^{+0.11}_{-0.10}$  & &  0.04$^{+0.02}_{-0.02}$ & \\
NGC 5253-4&	6 $^{+0}_{-2}$  && 1.62$^{+0.52}_{-0.48}$ & &  0.32$^{+0.04}_{-0.04}$ & \\
NGC 5253-5 &1 $^{+1}_{-1}$ &1$^{+1}_{-1}$& 7.46$^{+0.20}_{-0.27}$ & 32.96$^{+3.35}_{-2.10}$  & 0.46$^{+0.04}_{-0.04}$ & 0.46$^{+0.04}_{-0.02}$  \\
NGC 5253-11 &	1 $^{+1}_{-1}$  &&  25.5$^{+6.7}_{-4.2}$ &  & 0.46$^{+0.06}_{-0.02}$  &\\
NGC 5253-6&	10 $^{+2}_{-1}$&	10 $^{+2}_{-1}$& 3.24$^{+1.33}_{-0.94}$ & 3.24$^{+1.33}_{-0.94}$&	0.12$^{+0.04}_{-0.02}$ & 0.12$^{+0.04}_{-0.02}$ \\
NGC 5253-7	& 10 $^{+3}_{-1}$&10 $^{+3}_{-1}$ &1.15$^{+0.30}_{-0.56}$ &1.15$^{+0.30}_{-0.56}$&	0.05$^{+0.04}_{-0.03}$ &	0.05$^{+0.04}_{-0.03}$ \\
NGC 5253-8&	15 $^{+3}_{-3}$ & 15 $^{+3}_{-3}$&	2.88$^{+1.64}_{-0.84}$	&	2.88$^{+1.64}_{-0.84}$& 0.16$^{+0.04}_{-0.04}$ & 0.16$^{+0.04}_{-0.04}$ \\
NGC 5253-9&	10 $^{+2}_{-1}$	& 10 $^{+4}_{-2}$	&5.13$^{+2.12}_{-1.50}$ & 8.76$^{+1.93}_{-1.01}$ &	0.26$^{+0.06}_{-0.04}$ & 0.26$^{+0.04}_{-0.03}$ \\
NGC 5253-10&	10 $^{+5}_{-2}$&  &3.63$^{+3.22}_{-1.34}$ &&	0.26$^{+0.04}_{-0.04}$ &  \\
NGC 5253-12 & - & -	& - &	-  & - & - \\
						
    \hline
    \end{tabular}
    \begin{minipage}{15cm}~\\
 \footnotesize{
 $^{a}$ Average values weighted by mass, representative of the dominant stellar populations contained in the COS pointings. \\
 $^{b}$ Total stellar mass, representative of the stellar populations contained in the COS pointings.\\
  $^{c}$ Average values, representative of the stellar populations contained in the COS pointings. \\
 }
\end{minipage} 
\end{table*} 

\begin{table*}[!h] 
  \centering
    \caption{Sample properties from the literature as reported in the LEGUS catalog \citep{ada17,coo19,coo22} }\label{tab:legus_properties}
    \def\arraystretch{1.25}
     \begin{tabular}{l cc cccc }
    \hline
    \hline
   Target & Age$_{\rm phot}$ & $\langle$ Age$_{\rm phot}$ $\rangle$$^{a}$ & M$_{\rm phot}$ & M$_{\rm Total}$$^{b}$ & E(B-V)$_{\rm phot}$ & $\langle$E(B-V)$_{\rm phot}$$\rangle$ $^{c}$ \\
   & (Myr) & (Myr) & (10$^{4}$ M$_{\odot}$)& (10$^{4}$ M$_{\odot}$) &(mag)&(mag) \\
    \hline
    \hline
NGC 5253-1&	3 $^{+1}_{-0}$ & 10$^{+1}_{-0}$ & 1.55$^{+0.12}_{-0.17}$ & 5.33$^{+0.20}_{-0.26}$ & 0.04$^{+0.02}_{-0.00}$ & 0.07$^{+0.01}_{-0.01}$ \\
NGC 5253-2&	- &- & - & & - & \\
NGC 5253-3&	4 $^{+0}_{-0}$  &  & 0.58$^{+0.07}_{-0.02}$  & &  0.01$^{+0.03}_{-0.00}$ & \\
NGC 5253-4&	14 $^{+1}_{-1}$  && 3.20$^{+0.59}_{-0.75}$ & &  0.15$^{+0.02}_{-0.04}$ & \\
NGC 5253-5 &1 $^{+0}_{-0}$ &1$^{+0}_{-0}$& 6.21$^{+0.67}_{-0.43}$ & 8.10$^{+0.37}_{-0.26}$  & 0.43$^{+0.03}_{-0.02}$ & 0.54$^{+0.03}_{-0.03}$ \\
NGC 5253-11 &1 $^{+0}_{-0}$  &&  1.89$^{+0.30}_{-0.28}$ &  & 0.64$^{+0.05}_{-0.05}$  &\\
NGC 5253-6&	10 $^{+0}_{-0}$&	10 $^{+0}_{-0}$& 4.01$^{+0.48}_{-0.31}$ & 4.01$^{+0.48}_{-0.31}$&	0.02$^{+0.03}_{-0.00}$ & 0.02$^{+0.03}_{-0.00}$ \\
NGC 5253-7	& 15 $^{+0}_{-0}$&15 $^{+0}_{-0}$ &2.46$^{+0.30}_{-0.19}$ &2.46$^{+0.30}_{-0.19}$&	0.02$^{+0.03}_{-0.00}$ & 0.02$^{+0.03}_{-0.00}$ \\
NGC 5253-8&	10 $^{+10}_{-0}$ & 10 $^{+10}_{-0}$&	2.11$^{+0.34}_{-0.24}$	&	2.11$^{+0.34}_{-0.24}$& 0.05$^{+0.11}_{-0.03}$ & 0.05$^{+0.11}_{-0.03}$ \\
NGC 5253-9&	10 $^{+47}_{-6}$	& 12 $^{+24}_{-3}$	&4.85$^{+0.79}_{-0.54}$ & 13.14$^{+2.75}_{-2.68}$ &	0.22$^{+0.02}_{-0.03}$ & 0.26$^{+0.01}_{-0.11}$ \\
NGC 5253-10&	13 $^{+0}_{-0}$&  &8.29$^{+5.44}_{-5.34}$ &&	0.29$^{+0.02}_{-0.22}$ &  \\
NGC 5253-12 & 1 $^{+0}_{-0}$& 1 $^{+0}_{-0}$	& 0.93$^{+0.07}_{-0.10}$ &	0.93$^{+0.07}_{-0.10}$  & 0.30$^{+0.02}_{-0.03}$ & 0.30$^{+0.02}_{-0.03}$ \\
						
    \hline
    \end{tabular}
    \begin{minipage}{15cm}~\\
 \footnotesize{
 $^{a}$ Average values weighted by mass, representative of the dominant stellar populations contained in the COS pointings. \\
 $^{b}$ Total stellar mass, representative of the stellar populations contained in the COS pointings.\\
  $^{c}$ Average values, representative of the stellar populations contained in the COS pointings. \\ }
\end{minipage} 
\end{table*} 
 \begin{figure}[!h]
   	  \centerline{\includegraphics[scale=0.65]{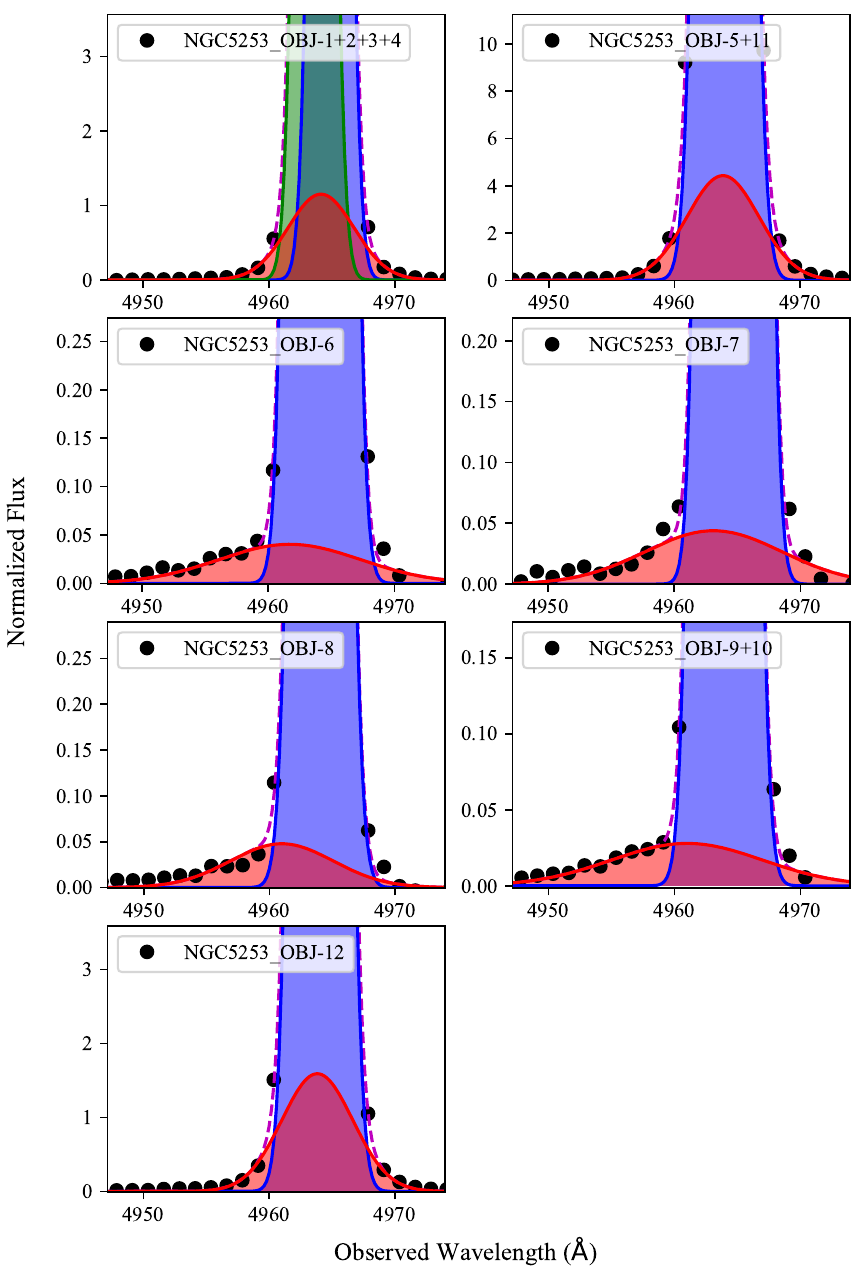}}
      \caption{Optical spectra from VLT/MUSE extractions corresponding to the same location of the COS pointings. The panels show a zoom-in around the [\ion{O}{3}] $\lambda$4959 forbidden line (black dots). The dashed purple line shows the best multi-component fits, and the shaded regions correspond to the different components (blue, red, green). }
         \label{fig:oiii_muse}
   \end{figure}

\subsection{Emission line analysis with VLT/MUSE}\label{sec:vlt}

\added{As part of the work presented in this paper, we investigate the possible influence of the targeted young massive clusters on their surrounding ISM through the analysis of optical IFS observations of NGC 5253 from VLT/MUSE previously presented in \citet{abr24}.} The MUSE observations were taken as part of program ID 095.B-0321 (PI. Vanzi, L.) and provide a wavelength coverage from 4750 to 9352 \r{A} at resolving power of $R\sim$ 3000. As described in \citet{abr24}, the data were reduced using the MUSE pipeline version v1.6 \citep{wei20}. Our work analyzes the extracted integrated spectra using circular apertures of 2.5\arcsec\ (12 spaxels) in diameter, centered on each of the COS pointings, including the additional NGC 5253-12 pointing. This approach allows us to infer the kinematic signatures of the ionized gas along the same line-of-sight as that of the YMCs studied in this work, probing comparable spatial scales (2\arcsec.5, $\sim$40 pc) in both the FUV and optical observations. Additionally, and especially relevant for this work, the ISM outflows probed by the optical [\ion{O}{3}] forbidden lines are expected to occur on relatively small scales ($\lesssim$100 pc), such as those of individual star clusters \citep{kom21, car25}. \par

We note that although previous kinematic studies of NGC 5253 have relied on the emission profile of H$\alpha$, for our work we use the [\ion{O}{3}] forbidden lines. \added{More specifically, we focus our analysis on the [\ion{O}{3}] $\lambda$4959 line over the brighter [\ion{O}{3}] $\lambda$5007 line, as the latter is blended with the weaker [\ion{He}{1}] $\lambda$5016 line.} Using this oxygen line also allowed us to avoid any line blending between the H$\alpha$ and [\ion{N}{2}] $\lambda\lambda$6548,6584 lines at the resolution of the MUSE observations. \added{Overall, the [\ion{O}{3}] $\lambda$4959 line was deemed to be ideally isolated from other emission providing us with a clean view of multiple kinematical components.} \par
We first normalized the spectral region of interest ($\lambda$ = 4776--5125 \r{A}) by fitting and subtracting the local continuum. Using tailored software we fit double and triple Gaussian profiles, and estimate the reduced-$\chi^2$ for each of the fits. Following the approach by \citet{wes09} and \citet{jam09}, we compared the reduced-$\chi^2$ of the two fits and apply a statistical $F$-test to determine which of the multi-component fits is most representative of the observed profiles. Under this statistical approach, the $F$-distribution function allows the calculation of the significance of a variance ($\chi^2$) increase associated with a confidence level depending on the assumed degrees of freedom. The test determines the minimum increase of the $\chi^2$ ratio between the two fits that is required at a given confidence limit to adopt a particular multi-component fit. Similar to recent emission-line studies \citep{jon25, her25}, we adopted a significance limit of 10\%, and assumed a two-component fit whenever its reduced-$\chi^2$ was lower by a factor of 3.289 than the estimated one for the three-component model. In Figure \ref{fig:oiii_muse} we show the adopted multi-component Gaussian fits. Overall, for the majority of the pointings we infer that the most representative model is the two-component profile, with the exception of NGC 5253-1+2+3+4. For this pointing, the $F$-test determines that a three-component fit is more appropriate. We note that this particular COS pointing encloses the largest number of individual clusters (Figure \ref{fig:pointings}), possibly responsible for a three-component Gaussian fit. \par

\section{Discussion}\label{sec:discussion}
\subsection{Characterization of young populations: Literature comparison}\label{sec:literature_stellar}
The clusters encompassed in the COS pointings have all been studied before in the literature. Most of these studies have focused on photometric analysis of HST observations \citep{cal97,har04,deg13}, with some of the most recent results reported in \citet[][C15 hereafter]{cal15a} and as part of the Legacy ExtraGalactic UV Survey \citep[LEGUS; ][]{cal15b, coo19, coo22}. Spectroscopic analyses characterizing the young clusters in NGC 5253 is more scarce. \citet{tre01} obtained FUV observations with the Space Telescope Imaging Spectrograph (STIS) on HST for one slit position covering the central star clusters, and derived ages of 1--10 Myr. Similarly, using these same STIS observations to study cluster NGC 5253-5, \citet{smith16} derived a very young age, $\sim$1-2 Myr.

Overall, these past studies report stellar ages younger than $\sim$15 Myr, confirming the youth of the starburst in the center of this dwarf galaxy as proposed in early studies \citep[e.g., ][]{van80, moo82}. The only exception comes from the photometric study by \citet{cal97}, where they estimate ages between 30 and 60 Myr for clusters NGC 5253-9 and NGC 5253-10. Studies have argued that the main reason for estimating such older stellar ages in such a young starbursting environment is the limited number of \added{broadband} filters, specifically the lack of filters close to the U-band limits \citep[critical for covering age-sensitive wavelengths;][]{lee05, cal15a}, which was indeed the case for the work by \citet{cal97}.\par
In the following subsections we compare our \texttt{SESAMME} results, listed in Table \ref{tab:sesamme}, against those inferred from these studies in the literature. 
 \begin{figure*}
   	  \centerline{\includegraphics[scale=0.65]{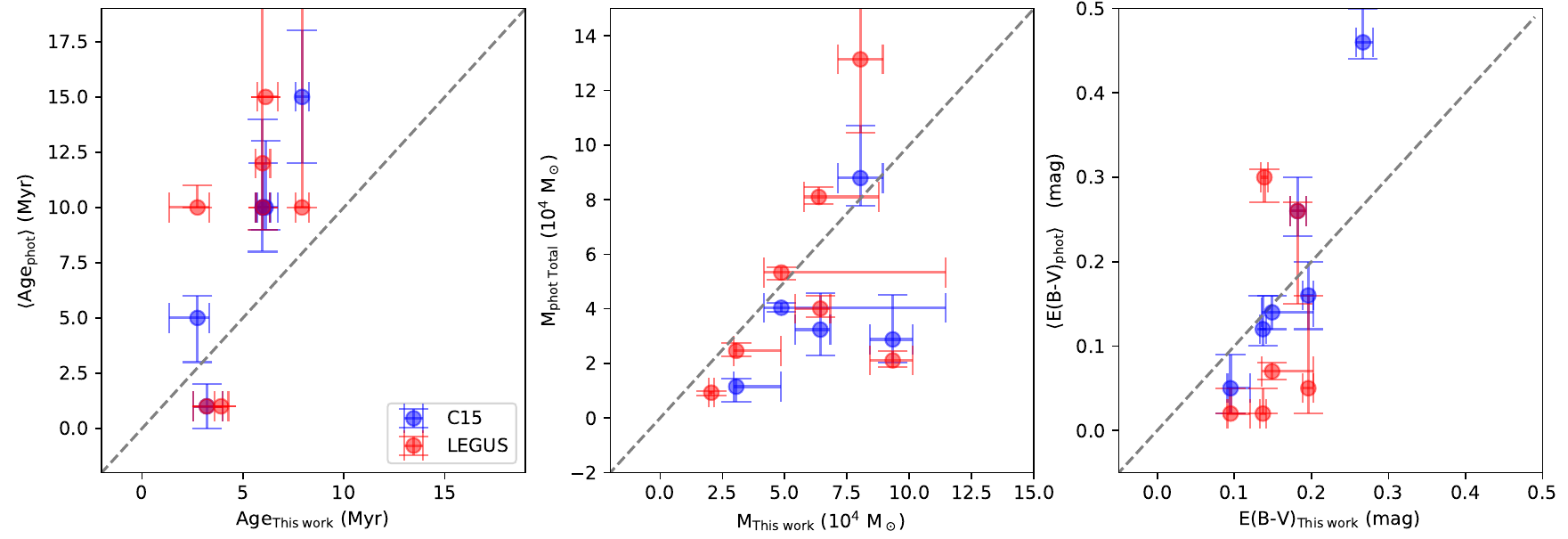}}
      \caption{Comparison between the properties inferred with \texttt{SESAMME} through the analysis of the COS FUV spectroscopic observations (on the x-axis), against the estimated values inferred through the photometric studies in the literature, C15 in blue, and LEGUS in red. The dashed-gray line shows the one-to-one values.}
         \label{fig:c15_comparison}
   \end{figure*}

\subsubsection{Age}
We first compare our spectroscopic ages with those reported by \citet{tre01}. 
The main aim of their study was to disentangle the diffuse UV emission from the light emitted by the stellar clusters. Their observations made use of the STIS 52\arcsec $\times$ 0\arcsec.1 long slit, covering stellar clusters NGC 5253-5, NGC 5253-9 and NGC 5253-10. One caveat with their analysis, specifically for NGC 5253-9 and NGC 5253-10, is the fact that the STIS slit does not fully encompasses the light from these two objects. According to \citet{tre01}, their STIS spectra capture only $\sim$6 \% and 3\% of the total luminosity of NGC 5253-9 and NGC 5253-10, respectively. They attribute this effect to the uncertainties in the exact position of the objects relative to the slit. 

Under the assumption that some of the light from these clusters was indeed captured by the STIS slit, \citet{tre01} estimate ages of 3 $\pm$0.9 Myr and 8 $^{+2.5}_{-0.9}$ Myr for NGC 5253-9 and NGC 5253-10, respectively, resulting in an average age of 5.5 $^{+1.3}_{-0.6}$ Myr. Assuming the analysis of the COS pointing covering these two clusters provides a resulting age estimate representative of the average ages of these two targets, the inferred age for NGC 5253-9+10 of 6.0$^{+0.4}_{-0.4}$ Myr agrees well with the reported spectroscopic ages by \citet{tre01}. 
\par
Cluster NGC 5253-5 has been studied spectroscopically by \citet{tre01} and \citet{smith16}.  \citet{tre01} report an age of 2 $^{+0.7}_{-0.8}$ Myr for this cluster, while \citet{smith16} find an age of 1--2 Myr assuming very massive stars are present. 
Using \texttt{SESAMME} we obtain an age of 3.2 $^{+0.8}_{-0.7}$ Myr for the COS pointing NGC 5253-5+11. This value is higher than the previously estimated spectroscopic age for NGC 5253-5 but we note that according to the LEGUS catalog, other clusters are encompassed in the $2''.5$ COS aperture, including the confirmed NGC 5253-11 cluster. \citet{smith20} show that the very central region of the starburst is complex and they find that NGC 5253-11 is very heavily extincted \citep{cal15a} and not visible in the FUV.
As a sanity check, and to further test the accuracy of \texttt{SESAMME} along with the \texttt{Starburst99} models, we perform a manual extraction of only NGC 5253-5 from the STIS spectrum. It is reassuring that the age of 1.8$^{+0.4}_{-0.4}$ Myr from this spectrum agrees with the originally reported value by \citet{tre01}, and it is within the range of ages estimated by \citet{smith16}.

\par

We now more generally compare the stellar properties inferred with \texttt{SESAMME}, through the analysis of the COS observations, against the latest photometric measurements for these same clusters as reported by C15 and in the LEGUS catalogs \citep{coo19,coo22}. One of the main differences between these two studies is the filters used for their SED fitting analysis. LEGUS made use of five HST filters (F275W, F336W, F435W, F555W, and F814W), whereas C15 relied on 13 different filters, including the F125LP filter covering the FUV band. 

In Tables \ref{tab:properties} and \ref{tab:legus_properties}, we list the properties of the individual star clusters as reported in C15 and in the LEGUS catalog, respectively, along with the average values, when applicable, representative of the UV bright stellar populations contained in the COS aperture. We note that the average ages reported in these tables are weighted by their respective masses to be more representative of the dominant stellar population. The average values estimated for these two studies are shown in Figure \ref{fig:c15_comparison}, as a function of our inferred spectroscopic estimates. We show with a dashed gray line the one-to-one line. From the left panel in Figure \ref{fig:c15_comparison}, we find that overall, the photometric ages by C15 (blue points) agree with our spectroscopic values within 2$\sigma$. The photometric ages from the LEGUS catalog (red points) show differences $>$2$\sigma$ compared to our \texttt{SESAMME} values. In general, the ages from both photometric studies are preferentially older than those estimated from the COS FUV spectra. \par
Past studies have extensively investigated and confirmed the existing degeneracies intrinsic to photometric studies of unresolved star clusters \citep[e.g.,][]{kav07, demeu13}. \citet{demeu14}, showed that stellar parameters are better constrained and suffer from weaker degeneracies when adding ultraviolet broadband photometry. We note that although C15 has included FUV photometric information in their SED analysis, degeneracies such as age-extinction may still be present, biasing the inferred properties. 

\subsubsection{Mass}
The middle panel in Figure \ref{fig:c15_comparison} compares the C15 and LEGUS stellar masses against the spectroscopic values inferred with \texttt{SESAMME}. Overall, the photometric masses by C15 tend to be consistently lower than what is obtained through the analysis of the FUV spectra, with the exception of the C15 mass estimate for NGC 5253-5+11, which has a M$_{\rm Total}$= 32.96$^{+3.35}_{-2.10} \times 10^4$ M$_{\odot}$ (off the limits of the middle panel in Figure \ref{fig:c15_comparison}). Aside from the NGC 5253-5+11 target, for the remaining five pointings for C15, we find that the inferred masses using the two different methods, photometric vs. spectroscopic, agree with each other only for two of those clusters, NGC 5253-9+10 and NGC 5253-1+2+3+4. The LEGUS photometric masses (red symbols in Figure \ref{fig:c15_comparison}) are in slightly better agreement with our spectroscopic COS measurements, with five pointings, NGC 5253-1+2+3+4, NGC 5253-5+11, NGC 5253-8, NGC 5253-9+10, NGC 5253-12, agreeing within 2$\sigma$. \par
A recent study by \citet{miz25} performed a comprehensive test of stellar population synthesis models via SED fitting (adopting aperture sizes that matched the COS aperture) and found that the agreement between broadband photometric SED fits and COS/FUV spectral fits is generally poor. Specifically on the mass estimates of star clusters, according to their study broadband photometry agrees with the FUV spectral fitting values only $\sim$20\%-60\% of the time. The differences we report here between our work and that by C15 and the LEGUS catalog are in agreement with the results from \citet{miz25}. As noted by \citet{miz25}, some of the differences observed between photometric and spectroscopic studies might also come from the application of different spectral synthesis models.

\subsubsection{E(B-V)}
Lastly, the right panel in Figure \ref{fig:c15_comparison} highlights the generally good agreement between the color excess inferred in the photometric study by C15 (blue points) and our spectroscopic study. We note that the most dramatic discrepancy between these two studies is observed in the YMCs with the highest E(B-V) value, NGC 5253-5+11. We note that C15 pointed out that cluster NGC 5253-11 has particularly high extinction, with large amounts of dust mixed with gas possibly biasing the overall E(B-V) measurements. Interestingly, the color excess estimates from the LEGUS catalog disagree with the values inferred with \texttt{SESAMME} even at the 2$\sigma$ level, with pointing NGC 5253-5+11 being the most discrepant with a $\langle$E(B-V)$_{\rm phot}$$\rangle$=0.54 off the limits of the right panel in Figure \ref{fig:c15_comparison}. One explanation for these discrepancies, and even the differences between the two photometric studies (LEGUS vs C15) is the more extensive wavelength coverage in C15, where they include the FUV band, helping to better constrain the reddening estimates. \par

\subsection{Stellar Feedback and Outflows}\label{sec:feedback_outflows}

Previous studies of NGC 5253 focusing on the gas kinematics of this starburst galaxy have reported complex flows in the central regions of the galaxy \citep[e.g., ][]{wal89, van06, lop07, coh18}. The complex kinematics in this system have been primarily uncovered through asymmetric emission line profiles with multiple components. Using VLT/FLAMES-ARGUS and Gemini-S/GMOS-IFU, \citet{mon10} and \citet{wes13}, respectively, found that the emission from H$\alpha$ across the main body of the galaxy was best fitted with two- and three-component profiles. Both studies attributed this complexity in the gas kinematics to winds from stellar clusters and multiple expanding shells. \par

\added{Based on the analysis of the optical MUSE observations, detailed in Section \ref{sec:vlt}, we now investigate the gas kinematics in the vicinity of the targeted YMCs. Given the blueshifted nature of the second component (shown in shaded red in Figure \ref{fig:oiii_muse}), we take this component to be representative of an outflow \citep[e.g., ][]{xu25}}. Figure \ref{fig:oiii_muse} clearly highlights a bi-modal distribution of profiles, specifically for the second component where four of the pointings require an extended profile (average FWHM $\sim$ 750 km s$^{-1}$, corrected for instrumental resolution), and three require a narrower fit (average FWHM $\sim$ 390 km s$^{-1}$). We note that for decades low surface brightness broad components were identified in nearby star-forming galaxies from both observational and theoretical studies \citep{cas90, gon94, ten97}. These components had typical FWHM of the order of $\sim$700 km s$^{-1}$, comparable to the observed features in NGC 5253. Spatially resolved studies such as that by \citet{wes09}, probing spatial scales of $\sim$10 pc, attribute this broad component to turbulent gas mixing \citep{sla93} which is believed to take place on the surfaces of interstellar clouds. The main driver of these turbulent layers is the impact of fast winds and high-energy photons originating from massive star clusters \citep[e.g.,][]{bin09}, a scenario that seems to apply to the environments probed in this work. \par

Our study exhibits a few trends that are worth noting: (1) when it comes to the second component, pointings with narrower profiles are all located in the northern region of the galaxy, NGC 5352-5+11, NGC 5253-1+2+3+4, and NGC 5253-12, (2) all three of these pointings contain the youngest star clusters studied here, with spectroscopic ages $<4$ Myr. These initial trends begin to hint at a connection between the energy injected into the ISM surrounding the YMCs, and the ages of these stellar populations. \par

\begin{figure*}
   	  \centerline{\includegraphics[scale=0.65]{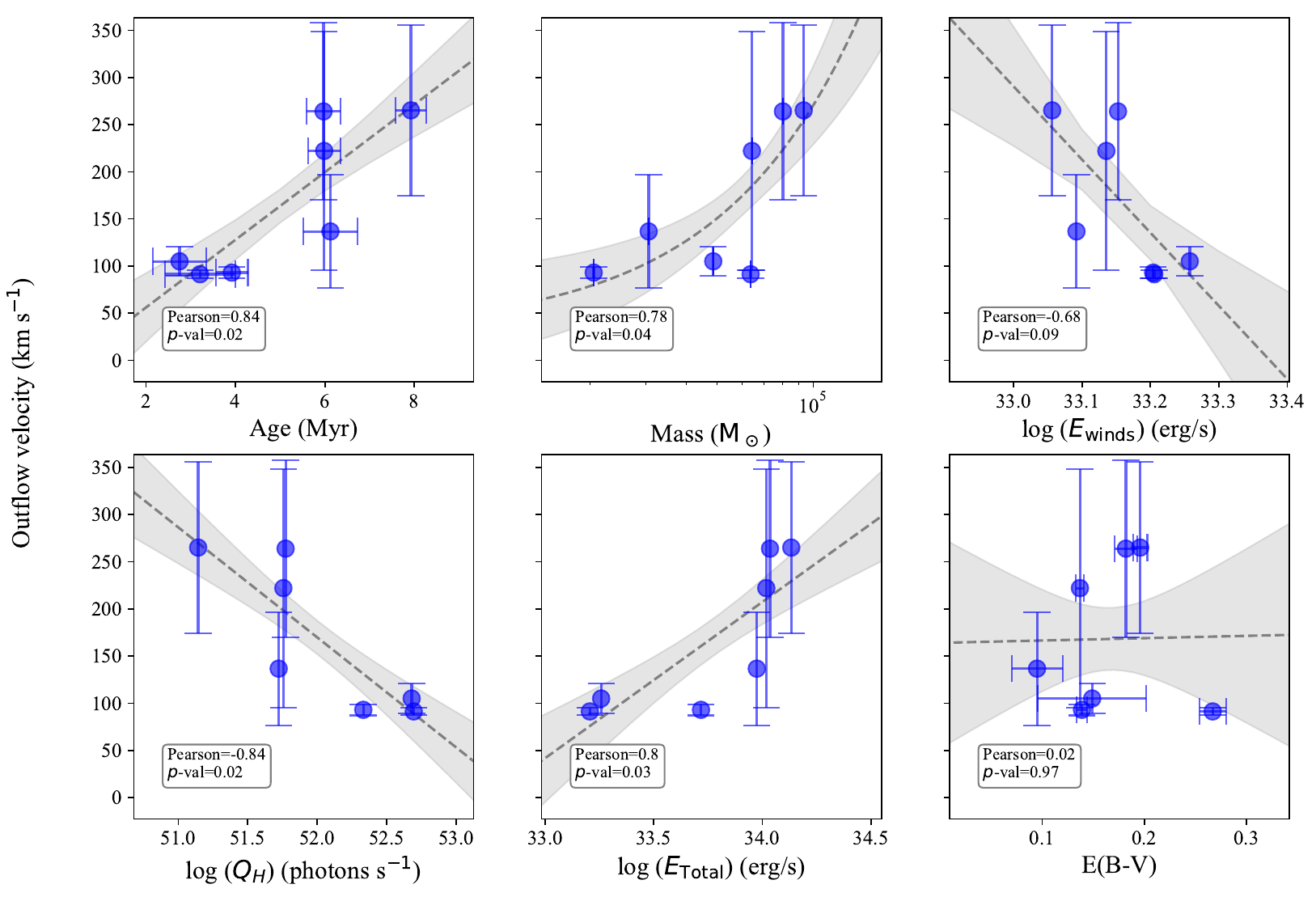}}
      \caption{Outflow velocities of the ionized gas (traced by [\ion{O}{3}] $\lambda$4959) as a function of the physical parameters inferred for the stellar populations targeted with COS. We show with a gray dashed line a linear regression. The legend in each panel shows the Pearson correlation coefficient and corresponding p-value. }
         \label{fig:outflows}
   \end{figure*}

The outflow velocities of the ionized gas at each pointing are estimated assuming a rest wavelength of 4958.9 \r{A} for the [\ion{O}{3}] line and a systemic velocity of 390 km s$^{-1}$ \citep{sch04} for NGC 5253. In Figure \ref{fig:outflows} we show in different panels the outflow velocities as a function of different stellar parameters as inferred with \texttt{SESAMME} and the \texttt{Starburst99} models. To assess the presence of a possible correlation between the outflow velocities and each of the physical parameters considered here, we apply a Pearson correlation test. We show in Figure \ref{fig:outflows} the linear regressions with gray dashed lines, and include the estimated Pearson correlation coefficients along with their p-values. We consider a statistically significant correlation in those parameters with p-values $<$ 0.05. \par

\subsubsection{Age} With a p-value of 0.02, we found a statistically significant correlation between the ionized gas outflows traced by [\ion{O}{3}] and the age of the YMCs (top left panel in Figure \ref{fig:outflows}). Although our sample is size-limited, we can see a clear change in the outflow velocities measured for YMCs with ages $<$ 5 Myr and those above this threshold. Theoretically, strong winds from stellar populations with ages $<$4-5 Myr are expected to dominate the feedback in star-forming systems \citep{lei95,lei99, gee23}. The latest population synthesis models by \citet{haw25}, predict that the peak of the mechanical energy from stellar winds occurs at stellar ages of 2-3 Myr with a clear dependence on metallicity and the presence of very massive stars (VMS; $\gtrsim$100 M$_{\odot}$). Our work would indicate that for the YMCs in our sample with ages $\leq$4 Myr, the observed outflows are primarily driven by their stellar winds. This is supported by the lack of non-thermal radio emission from the central starburst surrounding COS pointings NGC 5253-1+2+3+4, NGC 5253-5+11, and NGC 5253-12, showing a lack of SNe activity \citep{bec96}.
According to theoretical stellar evolution models and observational studies, at ages of $\gtrsim$4 Myr the first supernovae explode injecting an additional budget of mechanical energy into the ISM \citep{lei99, bik18}. This might be the cause of the observed increase in the outflow velocities for those clusters with ages $>$5 Myr, NGC 5253-6, NGC 5253-7, NGC 5253-8, and NGC 5253-9+10. \par

\citet{sir24} performed a comparable study, where they investigated the presence, or lack, of correlations between the outflow velocities of the ISM and their inferred stellar parameters, including the age of stellar clusters. When looking at the outflow velocities of the ionized phase of the atomic gas  (e.g., \ion{C}{4}, \ion{Si}{4} in absorption, after removing stellar contamination), their study found no correlation between this parameter and the stellar ages of their sample. One particular difference between the study presented here, and that by \citet{sir24} lies on the focus of the ISM component. Our work targets the ionized gas in the \ion{H}{2} regions surrounding the YMCs whereas \citet{sir24} probes the neutral gas reservoirs along the line of sight of the YMCs, which are located at larger distances from the stellar clusters than the localized ionized gas. In this context, the lack of correlations observed in the work by \citet{sir24} might be hinting at the extent of the feedback at play. This feedback is indeed impacting the gas in the vicinity (\ion{H}{2} regions), with the atomic/neutral ISM at larger distances appearing to be less influenced by these massive clusters. \par
For a more direct comparison between the work by \citet{sir24} and ours, we performed a similar analysis as theirs to quantify the outflow velocities of the more distant neutral gas. Details of this analysis are included in Appendix \ref{appendix:neutral}. Briefly, fitting multiple Voigt profiles to the \ion{S}{2} $\lambda$1250, 1253 transitions, we find a similar trend between the outflow velocities in this cooler ISM phase and the ages of the YMCs, Figure \ref{fig:neutral_veloffset}, as that observed in the ionized gas. One critical difference between what is observed in the ionized gas, compared to the trend in the neutral gas is the absolute outflow velocities, where in the latter we measure lower values. This difference is expected as the distance between the neutral gas reservoirs and the stellar clusters is much larger than that between the YMCs and the surrounding ionized gas. \par
Given that we observe a strong correlation between the neutral-gas outflow velocities and the ages of the YMCs in NGC 5253, and as this trend is not seen in the work of \citet{sir24}, we suggest that the nature of their sample may be driving this discrepancy. It is worth noting that in contrast to the study presented here where we characterize the stellar populations in a single low-metallicity galaxy, the analysis in \citet{sir24} included 1-2 YMCs across eleven different star-forming galaxies. The lack of correlation between ISM outflow velocities and stellar age in the work by \citet{sir24} might also be influenced by the variations in environments probed with their sample, which covered metallicities from sub-solar to extra-solar \citep[12+log(O/H)= 8.0–9.0;][]{sir22} across a diverse range of morphologies (grand-design arms, circumnuclear starburst rings, occulent spirals, and dwarf starbursts). Lastly, \citet{sir24} also highlighted that the lack of correlations in their study might be due to the large uncertainties in the reported outflow velocities of the ionized phase of the neutral gas which naturally resulted in large scatter of the data points. \par


\subsubsection{Mass} Similar to the trend with age, we find a positive statistically significant correlation (p-value = 0.04) between the outflow velocities of the ionized gas and the mass of the YMCs in NGC 5253. This correlation is present across the mass ranges estimated with \texttt{SESAMME}, from $\sim$ 10$^{4}$ to $\sim$ 10$^{5}$ M$_{\odot}$. It indicates that the higher the mass of the stellar populations, the higher the impact of their feedback into their surrounding ISM through energy and momentum. Once more, the YMC study by \citet{sir24} exhibited a lack of correlation between the outflow velocities of the high-ionization species in the atomic gas and their stellar mass. We note that as mentioned in the previous section, \citet{sir24} probe a wide range of environments and estimate masses in the range of $\sim$10$^{4}$ to $\sim$10$^{6}$ M$_{\odot}$, broader than the mass regime explored in NGC 5253 with this work. 

\subsubsection{Stellar Wind Mechanical Luminosity and Photoionization rate}\label{sec:stellar_wind}
Figure \ref{fig:outflows} shows an anticorrelation between the outflow velocities of the ionized ISM in the vicinity of the YMCs and both the theoretical stellar wind mechanical luminosity, log ($E_{\rm winds}$), and the theoretical photoionization rate, log ($Q_{\rm H}$). Both of these parameters are inferred through \texttt{SESAMME} and the models by \texttt{Starburst99}. We note, however, that between these two trends only the anticorrelation with photoionization rate is statistically significant (p-value = 0.02). For both of these trends the highest values of mechanical luminosity and photoionization rate are expected from the youngest clusters in the sample, ages $<$ 5 Myr. Overall, our work hints at a scenario where both the thermal radiation pressure by the ionizing photon flux and the mechanical luminosity from the winds of the most massive stars present are important components of feedback, driving the critical pre-SNe feedback \citep{che20, bar21, barr21}. \par
To this point, the recent work by \citet{pru25} which mapped nitrogen enrichment at 2.3-pc resolution, reported that the peak of the enrichment was located North of the clusters NGC 5253-5+11. Given that they found no direct spatial overlap between the nitrogen enriched ISM and the location of their confirmed Wolf-Rayet stars (WRs) and very massive stars \citep[VMS;][]{smith16}, they propose that indeed the enriched material has been expelled from their original sites through the energetic stellar winds of these massive stars. Together the observed correlations between gas outflows and stellar wind mechanical luminosity in our work, along with the recent results from \citet{pru25}, and other studies such as those by \citet{are25} and \citet{jam26}, underscore the role of massive stars in pre-SNe feedback.  \par

Our findings are also consistent with previous empirical and theoretical studies which have reported that both the thermal radiation pressure and stellar winds are critical pieces in the pre-SN feedback puzzle \citep{ben22, che22, del22, pat25}. Although theoretical studies have begun to explore feedback in the low-metallicity regime \citep[$\lesssim$0.4 Z$_{\odot}$;][]{jec23}, the findings reported here are relevant for setting observational constraints on the mechanisms that drive pre-SN feedback specifically in the unexplored metal-poor regime ($\lesssim$0.3 Z$_{\odot}$; Table \ref{tab:sesamme}). We note that even when our results support a scenario where both stellar winds and photoionization by massive stars play an essential role in injecting energy and momentum into the ISM, our work is unable to conclusively identify the most dominant mechanism in the stages before SNe occur. 

\subsubsection{Total Mechanical Luminosity: SNe contribution} 
We show in the lower middle panel in Figure \ref{fig:outflows} the observed trend between the ionized gas outflow velocities and the theoretical  total mechanical luminosity, log ($E_{\rm Total}$), which includes contributions from both stellar winds and SNe. Similar to log ($E_{\rm winds}$) and log ($Q_{\rm H}$), log ($E_{\rm Total}$) is inferred from the output parameters estimated by \texttt{SESAMME} and the \texttt{Starburst99} models. As the panel shows, we report a statistically significant strong correlation with a Pearson correlation coefficient value of 0.8 and p-value = 0.03. As discussed in Section \ref{sec:stellar_wind}, we detect an anticorrelation between the outflow velocities and the mechanical luminosity by stellar winds, which indicates that the statistically significant correlation observed between the outflow velocities and the total mechanical luminosity is primarily driven by the contribution from SNe. Interestingly, the four clusters enhancing this correlation are those with higher total mechanical luminosity, and therefore higher SNe mechanical luminosity. These four clusters are also those with the oldest estimated ages ($>$5 Myr; upper left panel in Figure \ref{fig:outflows}). This pattern is comparable to that observed by \citet{amo24}, and other studies in the literature \citep[e.g.,][]{bai24, flu25}, reporting that star-forming galaxies classified as weak Lyman continuum leakers (and therefore low log($Q_H$) values) are predominantly influenced by SNe-based feedback exhibiting outflows imprinted in collisionally excited emission lines (e.g., [\ion{O}{3}]), and dominated by older ($>$6 Myr) stellar populations.\par
Although these initial results propose a connection between the increased ionized ISM outflow velocity near clusters with ages $>$5 Myr and mechanical luminosity from SNe, we highlight that the latter parameter relies entirely, and is dependent, on theoretical models. Although previous studies of NGC 5253 such as that by \citet{tur98} have reported regions of diffuse non-thermal synchrotron emission possibly pointing to the presence of supernova remnants, a recent study by \citet{abr24} observed a lack of oxygen enrichment in the ionized gas phase, which in turn may be indicative of a lack of SNe events. An alternative scenario to explain the lack of co-spatial oxygen enrichment in the ISM from SNe in these YMCs is the possible expulsion of these enriched ejecta \citep{mcq15}. \par
In general, mechanical feedback, especially its initial phases, is expected to be decreased in low-metallicity environments compared to what is observed in more metal-enhanced systems \citep{lei14, ram19, vin22}. More specifically, at low metallicity SNe are predicted to start at later times than those expected for solar-metallicity environments. Models suggest that in metal-poor conditions a large fraction of potential progenitors of core-collapse SNe instead of exploding form black holes \citep{zha08, pat20}. According to \citet{jec23}, the injection of mechanical energy and momentum into the ISM by SNe is delayed in low-metallicity systems; instead of massive stars starting to explode at ages of $\sim$4-5 Myr, they are active until ages of $\sim$10 Myr in sub-solar environments (see their Figure 1). \par
In this context, our results where we find a clear increase in the outflowing velocities of the ionized gas surrounding those clusters with ages $>$5 Myr does not necessarily indicate or hint at a scenario with delayed mechanical feedback, in spite of probing relatively low metallicities ($\sim$0.2-0.3 Z$_{\odot}$). One possible explanation is that the decrease and suppression of mechanical feedback due to SNe explosions might be dominant at even lower metallicities. Interestingly, \citet{jec23} note that this change in feedback should begin to take place at metallicities $\lesssim$0.4 Z$_{\odot}$, although our \texttt{SESAMME} metallicity (stellar) estimates imply metallicities of Z $\sim$0.3 Z$_{\odot}$, literature abundances propose gas-phase metallicities for this galaxy as high as $\sim$0.37 Z$_{\odot}$ \citep{monreal12}, right at the metallicity limit proposed by \citet{jec23}. Lastly, in contrast to our findings, studying a region near our COS pointing NGC 5253-12 (for which we estimate a \texttt{SESAMME} age of $\sim$4 Myr), \citet{tur17} report that the CO gas surrounding the YMC in this region is relatively undisturbed indicating a lack of superwinds which in turn may hint at possibly suppressed stellar-wind feedback \citep{jec23}. \par

\subsubsection{E(B-V)} 
Given that part of the \texttt{SESAMME} analysis included estimates of the color excess, E(B-V), along the line of sight of the YMCs, we search for possible correlations between these values and the velocities of the outflowing ionized gas. Perhaps unsurprisingly, we find a lack of correlation between these two parameters (Pearson coefficient value of 0.02). \par

Overall, our work confirms the complex and intricate dependencies between the different physical parameters highlighting the fundamental role played by various feedback mechanisms dominating at different stages of star formation.

\section{Conclusions}\label{sec:sum}
We present the results from a UV study focused on the characterization of the young stellar populations in the blue compact dwarf, NGC 5253. The UV light from these YMCs is probed through a series of HST/COS pointings. We analyzed the FUV spectroscopic observations of seven pointings, and report on the physical parameters of the stellar populations encompassed in the 2\arcsec.5\ COS aperture as inferred by \texttt{SESAMME}. Here are our main findings: 
\begin{itemize}
\item We tested the sensitivity of the full spectral fitting code \texttt{SESAMME} to the wavelength coverage of the FUV observations ($\sim$1100-1775 \r{A} vs $\sim$1100-1425 \r{A}), and find that the broader wavelength coverage has a negligible effect on the best fitting age and metallicity. Limiting the wavelength coverage, however, appears to bias the inferred stellar masses and E(B-V), inferring higher masses by $<$50\% and higher E(B-V) values by $\sim$30\%. 
\item When comparing the stellar ages inferred from the spectroscopic analysis against values obtained through photometric studies such as those by \citet{cal15a} and the LEGUS survey, we find that the spectroscopic ages are preferentially younger (by $\sim$4 Myr) than those estimated through photometry. A possible reason for this might be the known degeneracies in photometric studies of unresolved star clusters \citep[e.g., ][]{kav07}.
\item Comparison between the photometric and spectroscopic masses showed agreement within the inferred uncertainties only for 20\% of the targets in the sample. Overall, the spectroscopic masses appeared to be consistently higher (by $\sim$1.4$\times$10$^{4}$ M$_{\odot}$) than what was estimated from the photometric study by \citet{cal15a}.
\item Overall, the color excess estimates inferred through the photometric analysis by C15 and our spectroscopic work agree within their uncertainties, with the highest discrepancy observed for the YMC with the highest E(B-V) value measured. The E(B-V) estimates obtained from the LEGUS catalog, on the other hand, disagree with our spectroscopic values at the level of $>$2$\sigma$. We believe the discrepancies between the results of these two photometric studies is driven by the number of filters used, primarily the inclusion of the bluest F125LP filter in the C15 study. 
\item Using VLT/MUSE optical observations and matching the size of the COS aperture, we measured outflow velocities of the ionized gas along the line of sight of each of the COS pointings, with values ranging from $\sim$125--300 km s$^{-1}$. 
\item We find strong statistically-significant correlations between the outflow velocities of the ionized gas around the observed YMCs and their inferred ages, masses, and total mechanical luminosity (primarily driven by SNe). \added{In particular, we find that the more massive and older ($>$4 Myr) clusters are associated with higher velocity outflowing gas, suggesting that their feedback has a higher impact on the ISM.} 
\item Given the clear correlation between outflow velocities and the total mechanical luminosity dominated by SNe, along with the statistically-significant trend between outflow velocities and the age of the clusters, we can conclude that even in this low metallicity ($\sim$0.2--0.3 Z$_{\odot}$) environment we confirm the absence of predicted delayed SNe feedback. 
\item We identify an anti-correlation between the outflow velocities and both the mechanical luminosity by stellar winds, and the photoionization rate. These trends hint at a picture where both of these mechanisms are critical drivers of feedback in the pre-SNe phase in this metal-poor environment. 
\end{itemize}

We note that studies similar to the one presented in this manuscript are critical to better understand the effect young massive clusters have on their surrounding ISM, particularly when pushing our understanding of metal-poor environments. The FUV regime provides a unique opportunity for fully characterizing young stellar populations and better constraining their physical properties. More importantly, the spatially-resolved nature of the work presented here highlights the incredible potential of future UV/Optical missions such as the \textit{Habitable Worlds Observatory}. The multi-wavelength nature of these types of work with FUV and optical spectroscopic coverage allow us to uncover the connection between the properties of the massive stellar populations and their impact to the nearby gas, shedding light into the critical role of the different feedback mechanisms shaping galaxy evolution.

\begin{acknowledgments}
SH and BLJ are thankful for support from the European Space Agency (ESA).
\end{acknowledgments}

%

\vspace{5mm}
\facilities{HST(COS), VLT(MUSE)}



\begin{figure*}
   	  \centerline{\includegraphics[scale=0.4]{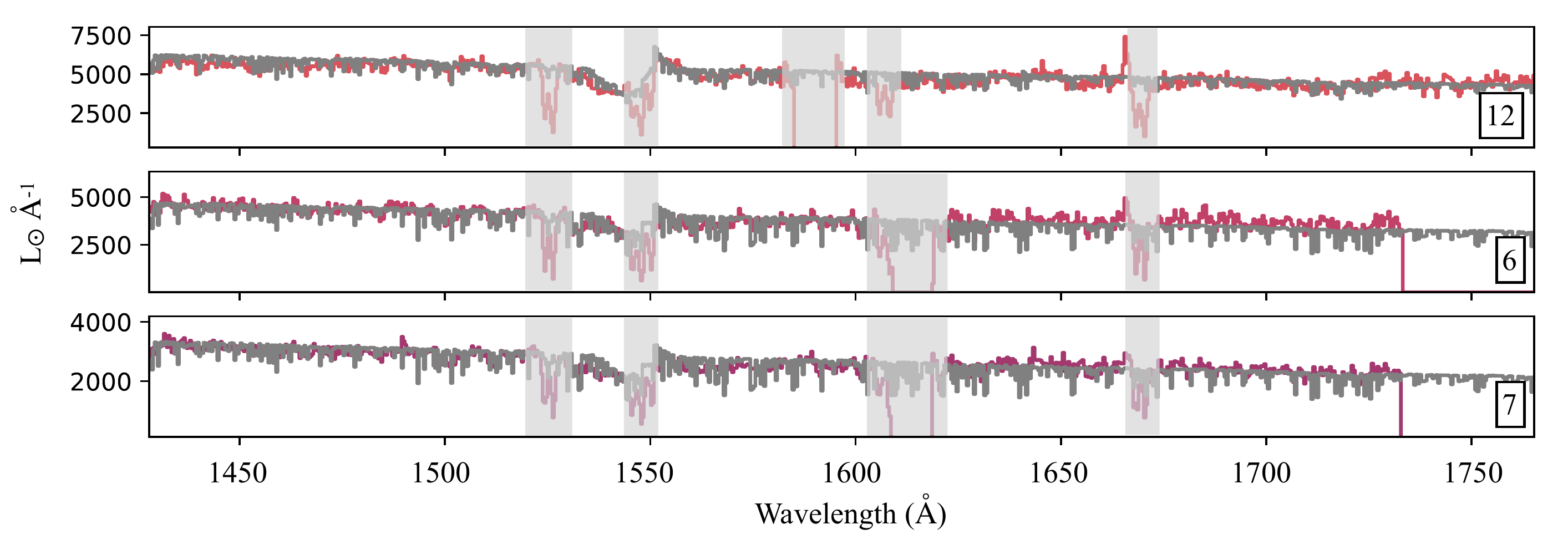}}
      \caption{Example synthesis fits for the three COS pointings with $\sim$1400-1775 \AA coverage. In gray we show the best stellar model identified by \texttt{SESAMME}. We note that the model does not include ISM contributions. The gray-shaded regions show the wavelengths masked in our full
spectral fitting analysis which primarily target strong absorption from the cold ISM.}
         \label{fig:fits_g160m}
   \end{figure*}

\appendix
\added{\section{Example synthesis fits}
We show in Figure \ref{fig:fits_g160m} the fits obtained by \texttt{SESAMME} for three of the clusters in our sample, NGC 5253-12, NGC 5253-6, NGC 5253-7. HST/COS FUV coverage between $\sim$1400-1775 \AA\; was available for all three of these targets. We detailed in Section \ref{sec:sesamme} the steps followed to characterize the dominant stellar population in each COS pointing. Overall, we note the general agreement between the model and observations for the \ion{C}{4} $\lambda\lambda$1548, 1550 wind lines.}

\begin{figure}
   	  \centerline{\includegraphics[scale=0.4]{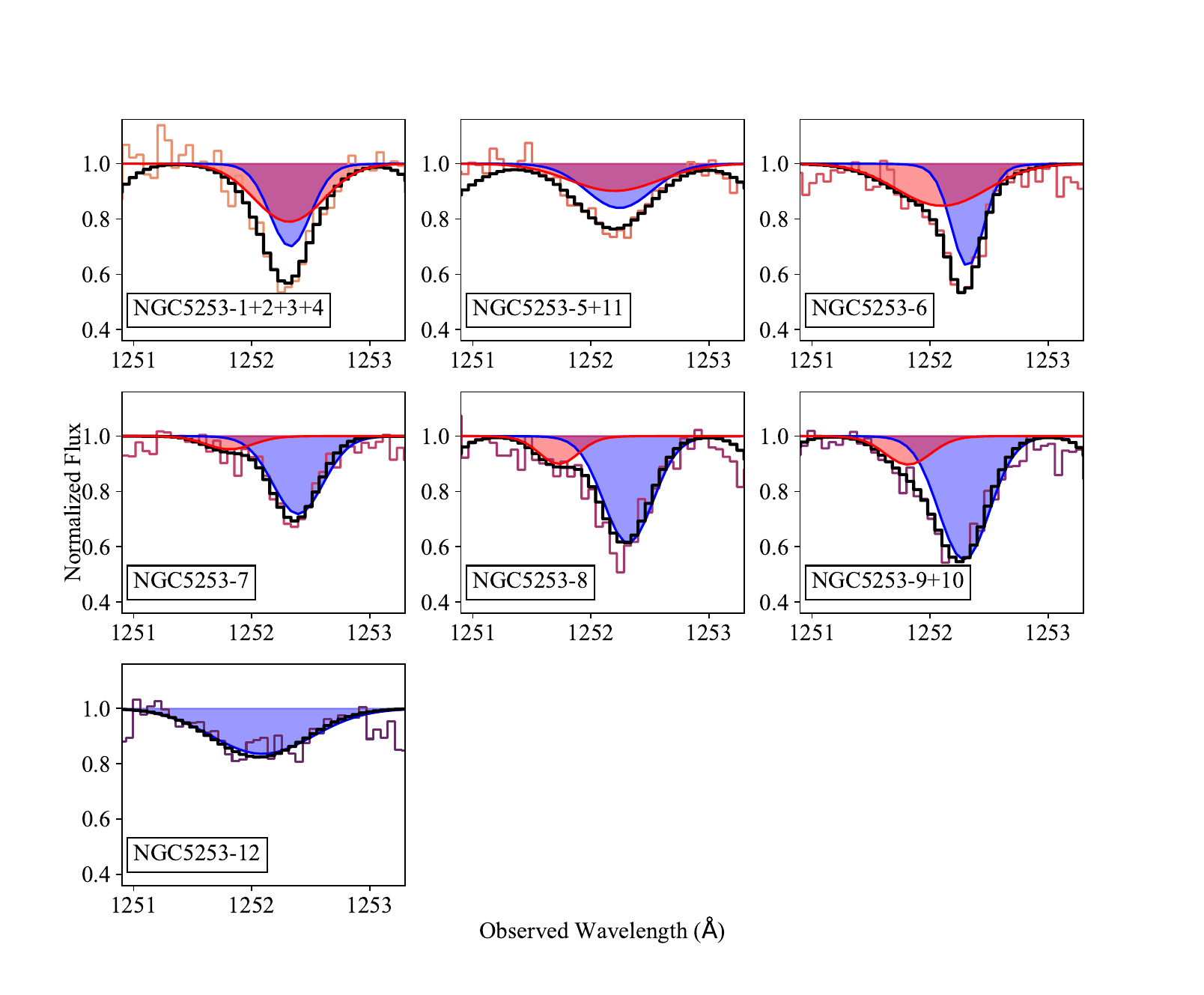}}
      \caption{Best profile fits for the \ion{S}{2} $\lambda$1250 transition. We show in blue the primary component, in red the secondary component, and the black solid line shows the profile with the combined components.}
         \label{fig:SII}
   \end{figure}

\begin{figure}
   	  \centerline{\includegraphics[scale=0.5]{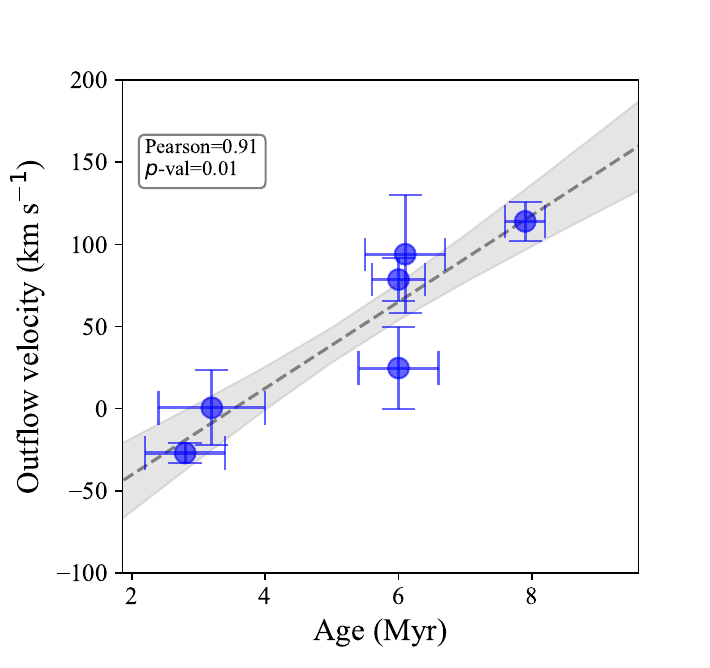}}
      \caption{Outflow velocities of the neutral gas (traced by the  \ion{S}{2} $\lambda$1250 transition) as a function of YMC age. We show with a gray dashed line a linear regression. The legend shows the Pearson correlation coefficient and corresponding p-value.} 
         \label{fig:neutral_veloffset}
   \end{figure}

\added{\section{Outflows in the neutral gas phase}\label{appendix:neutral}
We performed a similar exercise as that reported in Section \ref{sec:feedback_outflows} searching for potential outflows, but this time in the more distant neutral gas. Similar to past studies \citep[e.g.,][]{sir24, xu25}, we fit Voigt profiles to the absorption of the \ion{S}{2} $\lambda$1250, 1253 transitions to trace the kinematics of the cold neutral gas along the line of sight to the YMCs. In Figure \ref{fig:SII} we show the best fitting profiles as inferred with \texttt{VoigtFit} \citep{kro18} for the \ion{S}{2} $\lambda$1250 transition. We find that most pointings require a two-component fit, with the exception of NGC 5253-12. Overall, we find that similar to what was observed in the ionized gas surrounding the YMCs, our analysis indicates the presence of a second blue-shifted component for the older clusters in our sample, NGC 5253-6, NGC 5253-7, NGC 5253-8 and NGC 5253-9+10. Assuming the second components in the neutral-gas fits are representative of an outflow in this ISM phase, we estimate outflow velocities adopting a systemic velocity of 390 km s$^{-1}$ \citep{sch04} for NGC 5253. We show in Figure \ref{fig:neutral_veloffset} the inferred outflow velocities for the neutral gas in the direction of the YMCs, as a function of their inferred ages. We confirm a correlation between these two parameters through the application of a Pearson correlation test. We find a statistically significant correlation between the outflow velocities of the neutral gas and the ages of the YMCs (p-value =0.01). We note that although the outflow trends observed in the much cooler neutral gas resemble those traced in the ionized gas, a clear difference between the two is their outflow velocities. We observe lower outflow velocities in the neutral gas, compared to those reported for the ionized phase. This difference is expected given the proximity of the ionized gas to the sources of feedback (i.e., YMCs).}



\bibliography{sample631}{}
\bibliographystyle{aasjournalv7}



\end{document}